\documentclass[10pt,twocolumn,longbibliography,amsmath,amssymb,floatfix,superscriptaddress]{revtex4-2}

\usepackage{graphicx}
\usepackage{dcolumn}
\usepackage{bm}
\usepackage{color}
\usepackage{txfonts}
\usepackage{microtype}

\begin{document}

\author{Yan-Fei Li}\email{liyanfei@xjtu.edu.cn}	
\affiliation{Department of Nuclear Science and Technology, Xi'an Jiaotong University, Xi'an 710049, China}
\author{Yue-Yue Chen}\email{yueyuechen@shnu.edu.cn}
\affiliation{Department of Physics, Shanghai Normal University, Shanghai 200234, China}		
\author{Karen Z. Hatsagortsyan}
\author{Christoph H. Keitel}
\affiliation{Max-Planck-Institut f\"{u}r Kernphysik, Saupfercheckweg 1,
69117 Heidelberg, Germany}



\title{ Helicity transfer in strong laser fields via the electron anomalous magnetic moment}

\date{\today}

\begin{abstract}

Electron beam longitudinal polarization during the interaction with  counterpropagating circularly-polarized ultraintense laser pulses is investigated, while accounting for the anomalous magnetic moment of the electron. Although it is known that the helicity transfer from the laser photons to  the electron beam is suppressed in linear and nonlinear Compton scattering processes, we show that the helicity transfer nevertheless can happen via an intermediate step of the electron radiative transverse polarization, phase-matched with the driving field, followed up by spin rotation  into the longitudinal direction as induced by the anomalous magnetic moment of the electron. With spin-resolved QED Monte Carlo simulations, we demonstrate the consequent helicity transfer from laser photons to the 
 electron beam with a degree up to  10\%, along with an electron radial polarization  up to 65\% after multiple photon emissions in a femtosecond timescale. This effect is detectable with currently achievable laser facilities,  evidencing the role of the  leading QED vertex correction to the electron anomalous magnetic moment in the polarization dynamics in ultrastrong laser fields.



\end{abstract}

\maketitle

The development of modern ultraintense laser facilities, with a record intensity  already reaching
$10^{23}$ W/cm$^2$ \cite{Yoon2021},  brings
about new possibilities for testing predictions of strong-field quantum electrodynamics (QED) theory. The typical
field strength characterizing the strong-field QED regime is the Schwinger critical field  $E_S=1.3\times 10^{16}$~V/cm  (corresponding to the intensity of  $I_S=4.6 \times 10^{29}$W/cm$^2$) \cite{schwinger1951gauge,Piazza2012}.
While directly not attainable by lasers, the critical field can be achieved using the Lorentz boost of  ultrarelativistic electrons in a head-on collision geometry \cite{jackson1999},  which
enables the experimental study of nonlinear regimes of strong-field QED processes. In particular, recently such experiments are proposed at DESY (LUXE) \cite{abramowicz2019letter}, and at FACET-II (E320) in the Stanford Linear Accelerator Center (SLAC) \cite{unpublished}.

In strong background fields electrons can be polarized due to the spin-flip during photon emissions, which was first discovered for synchrotron radiation \cite{Sokolov1964,baier1971radiative,Baier1972,Derbenev_1973,Baier1998} and termed as radiative polarization.  Recently, the possibility has been proven of efficient radiative polarization using ultrastrong laser fields, applied to produce polarized electrons \cite{Liyf2019,Seipt2019} and positrons \cite{Liyfei2020,Chen2019,Wan2020} in a femtosecond timescale, as well as for polarization transfer from electrons to $\gamma$-rays in laser fields \cite{Liyf2020,Tang2020}.
The polarization effects in strong laser fields have a capability of detecting the quantum stochastic nature of electron dynamics \cite{Guo2020},  diagnosing  magnetic fields of plasma \cite{Gong_2021}, and providing ultra-short, high-brilliance, low-emittance polarized beam sources for fundamental studies in high-energy physics
\cite{Moortgat2008,Subashiev1998,Elouadrhiri2009} and material science \cite{Rich1987,Gidley1982}.  The completely spin- and photon-polarization-resolved probability rates for nonlinear Compton scattering have been derived from
strong-field QED theory in the Furry picture  for a plane-wave laser field \cite{Ivanov_2004}, and for the locally constant fields \cite{Seipt2020}, as well as via the quantum operator method \cite{wistisen2019numerical}, and employed  for a deep analysis of all polarization channels, helicity transfer in the perturbative regime,  and investigation of the polarization dependent energy and angle distributions \cite{king2020nonlinear}.  Furthermore,  the study of polarization effects has been extended to higher-order QED processes \cite{Baier_1975,Meuren2011,podszus2021first, Ilderton2020,Torgrimsson_2021,torgrimsson2021resummation}, and QED cascades  \cite{Seipt2021}.

In radiative polarization  the electron spin-flip is preferable along the instantaneous magnetic field in the rest-frame of the electron. Because of that in a storage ring or in ultrastrong laser fields  initially unpolarized electrons are mostly polarized transversely after the interaction. Nevertheless, high-precision high-energy physics at accelerators demands  longitudinal beam polarization, e.g., the Q$_{\rm{weak}}$ experiment at Jefferson Lab \cite{carlini2012qweak} and E158 at  SLAC \cite{kowalski2020parity}.  Presently, one common way of  producing circularly polarized electrons in accelerators is via photoemission  induced by a circular polarized laser field from a solid \cite{Pierce_1976,Swartz_1988}. There were a series of attempts to employ Compton scattering for this purpose. While here the helicity of the laser photons is efficiently transferred  to emitted $\gamma$-rays \cite{Ohgaki_1996,Omori_2006}, the  longitudinal polarization of the scattered electrons is not efficient in the linear regime and suppressed in the nonlinear regime. Thus, the electron longitudinal polarization can reach only  $P_\parallel\sim 10^{-3}$ during a single photon emission in a circularly polarized laser field \cite{Seipt2018} at the laser strong field parameter $a_0=100$. Moreover, while scattered electrons in the Compton process are weakly polarized, the total longitudinal polarization of the electron beam is vanishing. This is due to the polarization of unscattered electrons, which exactly cancels that of scattered ones \cite{Kotkin2003,Karlovets2011}.  The latter is explained as interference of the incoming electron wave function with that of the forward scattered one.

Recently, it has been demonstrated that QED radiative corrections, i.e., the interaction of the electron with its own radiation field, can also affect the electron spin dynamics in intense background fields \cite{Meuren2011,podszus2021first,Ilderton2020}.  In particular, due to QED loop corrections, the electrons exact spin-dependent wave function becomes unstable inside a strong background field, leading to 1\% polarization for unscattered electrons \cite{Meuren2011,podszus2021first}. The latter provides the QED description of the polarization of unscattered electrons discussed in \cite{Kotkin2003}. The spin effects resulted from the electron mass loop are also described as a spin rotation, which appeared to be more significant within the tail of a tightly focused laser beam \cite{Ilderton2020}.
Furthermore, there are experimental plans to reach the fully nonperturbative regime of QED employing beam-beam collisions in TeV-class lepton colliders \cite{di2020testing,blackburn2019reaching,baumann2019probing,fedotov2017conjecture,yakimenko2019prospect,Mironov2020}, when the effective field in the rest frame of electrons could be supercritical, and  radiative corrections to QED processes nonperturbative and substantial.

In this Letter, we investigate the role of the electrons anomalous magnetic moment on the helicity transfer from a circularly polarized (CP) laser pulse to an ultrarelativistic electron beam for the nonlinear Compton scattering  process in the radiation reaction dominated regime. The electron three-dimensional polarization properties are analyzed using numerical Monte Carlo simulations based on the spin-resolved radiation probabilities in the local constant field approximation (LCFA).  While previous studies neglecting the QED radiative corrections came to a conclusion that the helicity transfer from laser photons to electrons is forbidden in the nonlinear Compton scattering process,we obtain  a sizable longitudinal polarization of electrons when the one-loop QED vertex correction \cite{ritus1970radiative,baier1971radiative} to the anomalous contribution to the magnetic moment is accounted for. A longitudinal  polarization degree close to $3\%$ is shown, which could be further improved up to 10\% with  post-selection techniques. We prove the scenario of the helicity transfer. Initially, spin-flips during photon emissions induce electron  transverse  polarization which is phase-matched with the laser field.  Due to the latter property and the anomalous correction to the magnetic moment \cite{Hanneke2008,Sturm2011,Abi2021}, the oscillating transverse  polarization is transformed into accumulated  longitudinal polarization during the interaction. The latter demonstrates a signature of QED radiative corrections for electron polarization dynamics in ultrastrong laser fields. Additionally, the electron  transverse  radiative polarization with a degree over 60\% is shown after the interaction, of interest for high-energy applications.


We model the laser-electron interaction process with the Monte Carlo method \cite{Ridgers2014,Elkina2011,Gonoskov2015,CAIN}, which treats the photon emissions quantum mechanically with the spin-resolved photon emission probabilities in the LCFA  \cite{Liyf2019,Liyf2020}. The LCFA \cite{Ritus1985,Baier1998,Piazza2018,Piazza2019,Ilderton2019,lv2021anomalous} is valid at $a_0\equiv|e|E_0/(m\omega_0)\gg1$, when the coherence length of the photon emission, $l\sim \lambda_L/a_0$, is much smaller than the typical length of the trajectory (here the laser wavelength $ \lambda_L$).
Furthermore, $E_0$ is the laser field amplitude, $\omega_0$ the laser frequency, and $e (<0)$, $m$ are the electron charge and mass, respectively. Relativistic units $\hbar =c=1$ are used throughout.
The photon emission probability is determined by the local value of the quantum strong-field parameter $\chi_e\equiv |e| \sqrt{-(F_{\mu \nu}p^\nu)^2}/m^3$, where $F_{\mu \nu}$ is the field tensor and $p^\nu$ the four-vector of electron momentum. The simulation method is the following \cite{SM}: The common statistical  event generator is conducted at each
simulation step to determine whether or not a photon-emission occurs. If a photon-emission occurs, the emitted photon energy is determined via the stochastic procedure and spectral probability, while the electron and  photon polarizations via the averaged algorithm involving the density matrix for the mixed state of an electron ensemble
\cite{CAIN,Tang2021} to reduce the statistical fluctuation. If the photon-emission event is rejected, the electron spin changes according to the non-radiation probability \cite{CAIN,Liyfei2020}. This no-emission spin variation originates from the radiative correction of the one-loop propagator correction where the electron propagator is modified by the process that a virtual photon is emitted and reabsorbed by the electron \cite{Meuren2011}.
At first order in the fine-structure constant $\alpha$,
it cancels  the longitudinal  polarization of the electrons induced by emitting soft photons in a circularly polarized field
{  \cite{Kotkin2003,kotkin1998polarization,Torgrimsson_2021}.
 In the regime $a_0\sim1$, the cancellation is broken after including radiation reaction, leading to a nonzero longitudinal polarization of the final electrons \cite{torgrimsson2021resummation}.

The spin precession between photon emissions is governed by the Thomas-Bargmann-Michel-Telegdi equation \cite{Bargmann1959}: ${\rm d}{\bf S}/{\rm d}t=\bm{S}\times\bm{F}$,
with
\begin{eqnarray}\label{F}
{\bm F}&=&\frac{e}{m}\left[-\left(\frac{g}{2}-1\right)\frac{\gamma}{\gamma+1}\left({\bm \beta}\cdot{\bf B}\right){\bm \beta}\right.\nonumber\\
&&\left.+\left(\frac{g}{2}-1+\frac{1}{\gamma}\right){\bf B}-\left(\frac{g}{2}-\frac{\gamma}{\gamma+1}\right){\bm \beta}\times{\bf E}\right],
\end{eqnarray}
where ${\bf E}$ and  ${\bf B}$ are the  laser electric and magnetic fields, respectively, $\beta \approx -1$ the electron velocity, and $g$ the  electron gyromagnetic factor. Taking into account the radiative correction to the first order of $\alpha$ in the interaction with the radiation field and being exact with respect to
the external field \cite{Baier1972},
\begin{equation}\label{ggg}
  g\left(\chi_e\right)=2+2\mu\left(\chi_e\right), \,\,\,\,\,\mu\left(\chi_e\right)=\frac{\alpha}{\pi\chi_e}\int_{0}^{\infty}\frac{y}{\left(1+y\right)^3}{\bf L}_{\frac{1}{3}} \left(\frac{2y}{3\chi_e}\right){\rm d}y,
\end{equation}
with ${\bf L}_{\frac{1}{3}} \left(z\right)=\int_{0}^{\infty}{\rm sin}\left[\frac{3z}{2}\left(x+\frac{x^3}{3}\right)\right]{\rm d}x$. At $\chi_e\ll1$, one obtains the Schwinger result  $g=2+\frac{\alpha}{\pi}\approx2.00232$.
 The electron dynamics is described by the Newton equation with the Lorentz force. The modification of the equation of motion due to the anomalous magnetic moment \cite{Wen_2016,Formanek_2021} does not change the electron dynamics \cite{SM}.

  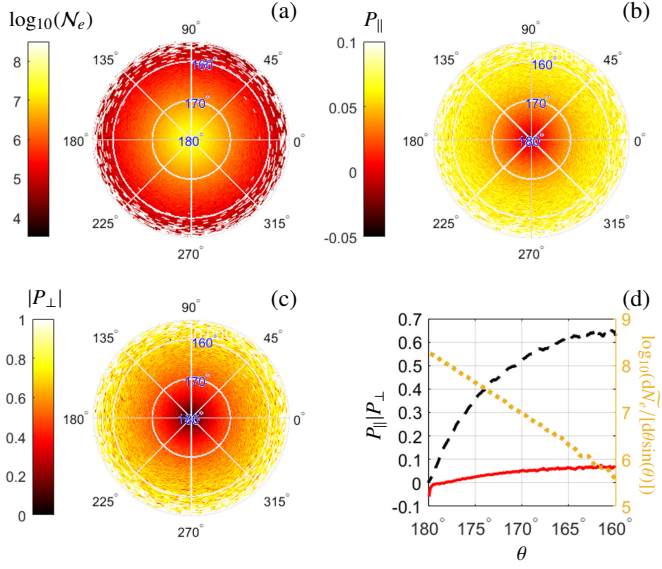
\begin{figure}[t]
 	\includegraphics[width=1\linewidth]{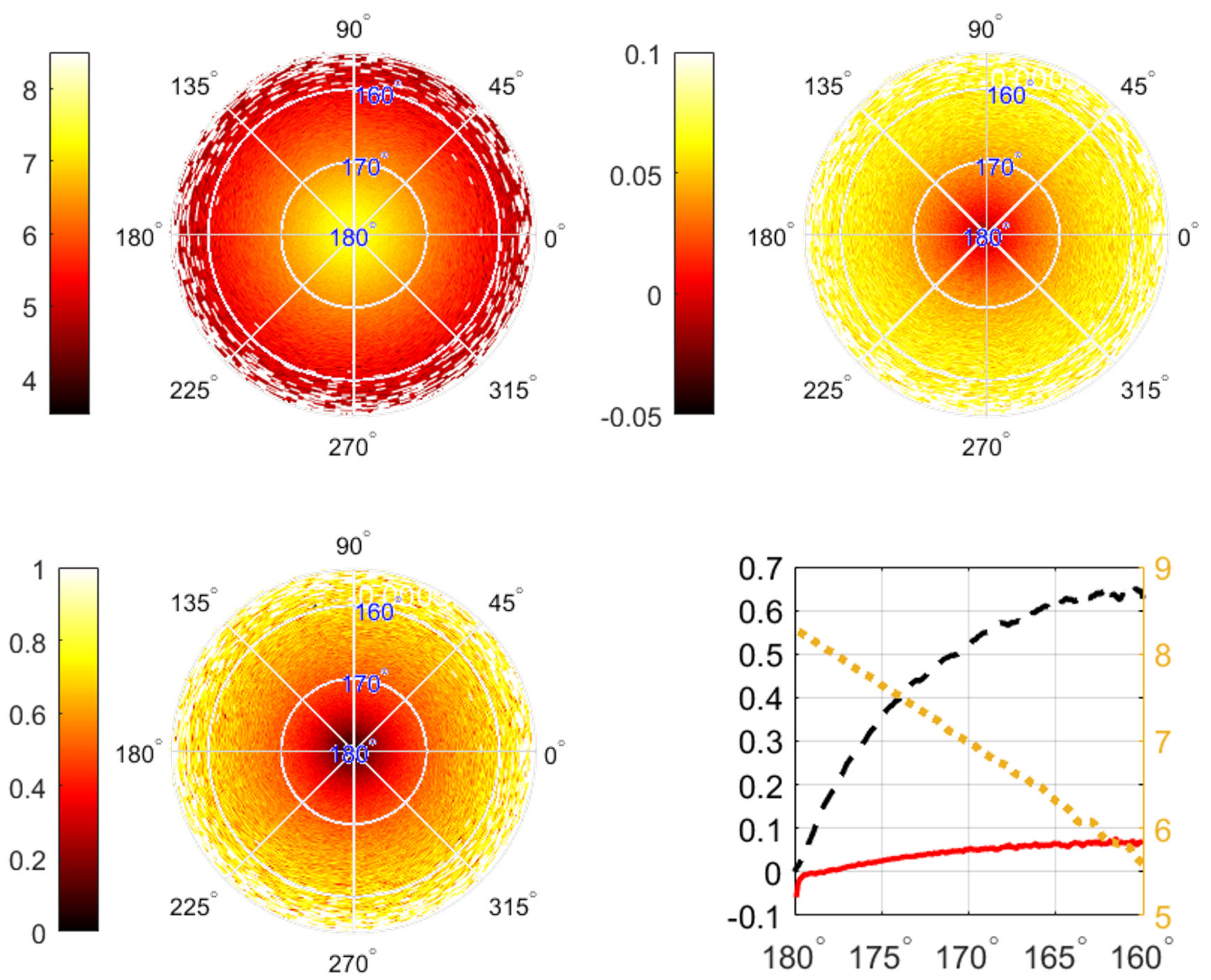}
 \begin{picture}(300,0)
  \put(102,208){(a)}
    \put(3,205){\footnotesize log$_{10}$($\mathcal{N}_e$)}

  \put(235,208){(b)}
     \put(137,205){\footnotesize $P_\parallel$}

   \put(102,99){(c)}
   \put(10,100){\footnotesize $|P_\perp|$}

    \put(235,99){(d)}
	 \put(241,88){\rotatebox{-90}{{\color[rgb]{0.93,0.69,0.13}\scriptsize log$_{10}($d$\widetilde{N_e}/[$d$\theta$sin$(\theta)$])}}	}
	 	 \put(138,48){\rotatebox{90}{\footnotesize $P_\parallel | P_\perp$}}	
	     \put(195,3){\footnotesize $\theta$}
\end{picture}	
 \caption{(a) Distribution of electron number density log$_{10}$($\mathcal{N}_e$), with $\mathcal{N}_e=$ d$N_e/[$d$\varphi$d$\sin\theta$ d$\theta$ d$\phi $]; (b) Average  longitudinal polarization $P_\parallel={\bm \beta} \cdot {\bm S_f}$; (c) Average transverse polarization $|P_\perp|=|{\bm S}_f-{\bm \beta} \cdot {\bm S_f}|$, vs polar angle $\theta$ in [$155^\circ, 180^\circ$] and the azimuthal angle of $\varphi$ in [$0^\circ, 360^\circ$], respectively; (d) $P_\parallel$ (red-solid), $P_\perp$ (black-dashed) and log$_{10}($d${N_e}/[$d$\theta$sin$(\theta)$]) (yellow-dotted) at $\varphi=180^\circ$ vs  $\theta$, respectively.} \label{fig1}
 \end{figure}

The polarization effect of electrons is illustrated in  Fig.~\ref{fig1}.   A right-hand CP tightly-focused Gaussian laser pulse is used, with peak  intensity  $I_0\approx 2\times 10^{22}$ W/cm$^2$ $ (a_0=100\sqrt{2})$, pulse duration (FWHM) $\tau=5T_0$, with the laser period $T_0$,  $\lambda=1 \mu$m, and focal radius $w_0=5 \lambda$. The counterpropagating cylindrical electron bunch has a length of $L_e=5 \lambda$ and radius of $w_e=1 \lambda$. $N_e=10^6$ unpolarized electrons are distributed longitudinally uniformly. The transverse distribution is Gaussian with the variance of $\sigma_{x,y}=0.3 \lambda$. The initial electron kinetic energy is  $\varepsilon_0=1$ GeV, the energy spread $\Delta \varepsilon_0/\varepsilon_0=10\%$, and the angular divergence (FWHM) $\Delta \theta=0.1$ mrad. The feasibility of our scheme for larger beam spreadings is shown in \cite{SM}. As the quantum parameter for pair creation $\overline{\chi}_{\gamma}\equiv |e| \sqrt{-(F_{\mu \nu}k^\nu)^2}/m^3 \approx 0.005\ll1$, with $k^\mu=(\omega,{\bm k})$ being the four-momentum of laser photons, the pair creation effect can be neglected \cite{SM}.

After the interaction the scattered electrons concentrate in the center of  the angle distribution, and have longitudinal, as well as transverse polarization, see Figs.~\ref{fig1}(a)-(c). The average longitudinal polarization degree is not large $\overline{P}_\parallel=\frac{1}{N_e}\sum\limits_{i=1}^{N_e}{\bm S_i}\cdot\frac{\bm \beta_i}{|\bm \beta_i|}\approx$ 2.65\%, nevertheless exceeding by an order of magnitude the QED tree-level result $\overline{P}_\parallel\approx 0.1\%$ \cite{Seipt2018}. The electrons are
highly polarized in the transverse plane with  polarization vector pointing to the center of the beam.
The electron number-density decreases exponentially from the center to the peripheries [Fig.\ref{fig1}(d)], while the polarization $P_\parallel$ and $P_\perp$ increase with the deflection angle.
Therefore, the polarization purity can be increased by selecting large angle electrons with post-momentum-angle selection techniques.
Meanwhile, the longitudinal polarization degree can be increased by post-energy-selection due to the $P_\parallel$ dependency on the electrons energy [Fig.~\ref{fig2}]. Higher $P_\parallel$  can be obtained by collecting low energy electrons. For instance, by collecting the electrons with energies less than 45 MeV,  we can get a polarization degree of  $P_\parallel=10\%$ with the percentage of number of 1\%.

  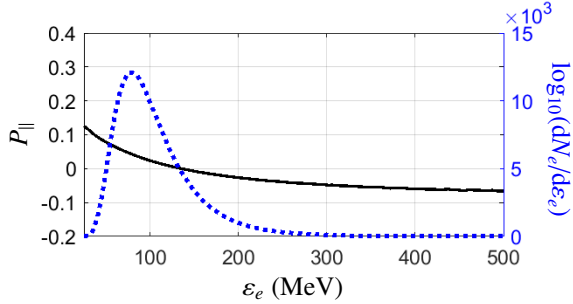
\begin{figure}[t]
 	\includegraphics[width=0.8\linewidth]{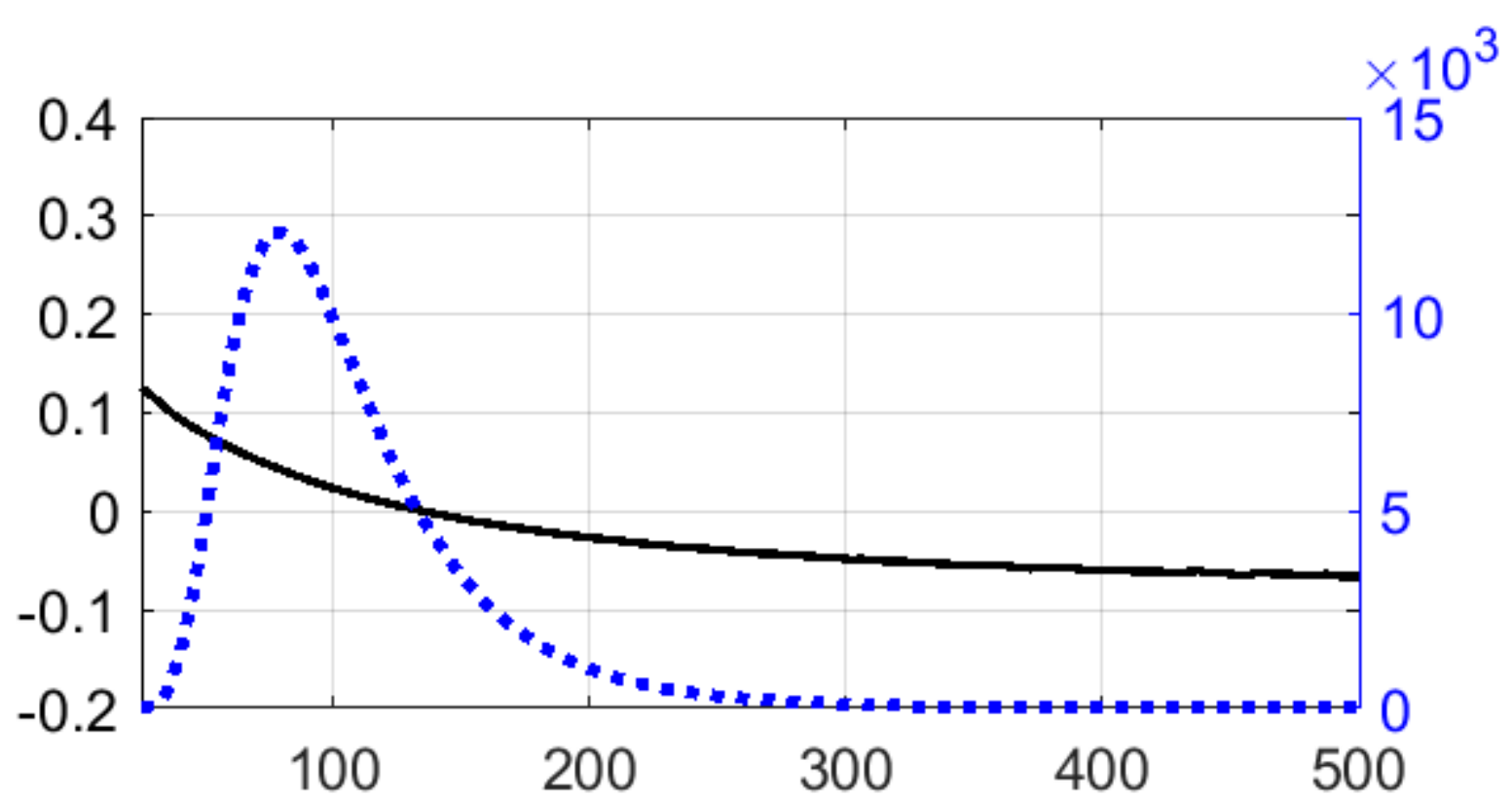}
 \begin{picture}(300,0)
 \put(15,56){\rotatebox{90}{\normalsize $P_\parallel$}}	
	 \put(218,88){\rotatebox{-90}{{\color{blue}\normalsize log$_{10}$(d$N_{e}$/d$\varepsilon_{e}$)}}	}
	 \put(103,0){{\normalsize $\varepsilon_e$ (MeV)}}	
\end{picture}
		 \caption{Longitudinal polarization $P_\parallel$, and number density log$_{10}$(d$N_{e}$/d$\varepsilon_{e}$) (MeV$^{-1}$), versus final electron energy $\varepsilon_{e}$, respectively. } \label{fig2}
 \end{figure}

 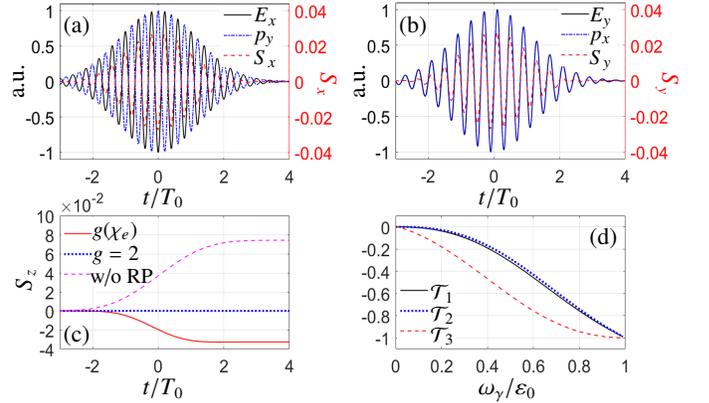
\begin{figure}[b]
 	\includegraphics[width=1\linewidth]{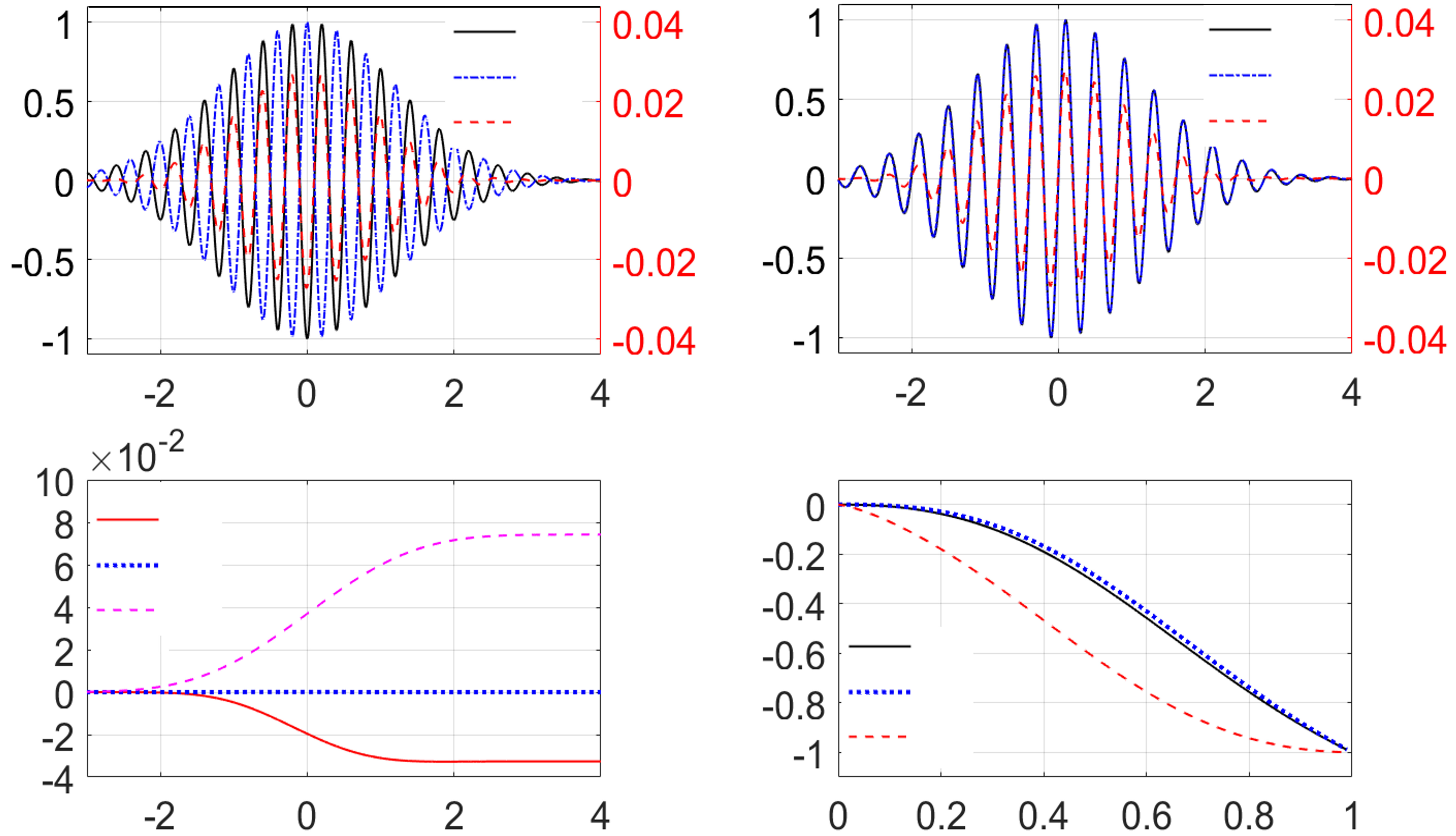}
 \begin{picture}(300,0)
  \put(16,140){\small (a)}
 \put(-3,115){\rotatebox{90}{\small a.u.}}	
 	 \put(46,75){{\small $t/T_0$}}	
 \put(112,125){\rotatebox{-90}{\small {\color{red} $S_x$}}}
 \put(88,144){{\footnotesize $E_x$}}	
 \put(88,137){{\footnotesize $p_y$}}	
  \put(88,128){{\footnotesize $S_x$}}	
 	
  \put(143,140){\small (b)}
	 \put(126,115){\rotatebox{90}{\small a.u.}}	
	 \put(173,75){{\small $t/T_0$}}
 \put(240,125){\rotatebox{-90}{\small {\color{red} $S_y$}}}
 \put(215,144){{\footnotesize $E_y$}}	
 \put(215,137){{\footnotesize $p_x$}}	
  \put(215,128){{\footnotesize $S_y$}}

  \put(16,23){\small (c)}
	 \put(-2,42){\rotatebox{90}{\small $S_z$}}	
	 \put(46,3){{\small $t/T_0$}}
    \put(27,63){{\footnotesize $g(\chi_e)$}}	
  \put(27,54){{\footnotesize $g=2$}}
  \put(26,45){{\footnotesize w/o RP}}
	
  \put(215,60){\small (d)}
	 \put(173,3){{\small $\omega_\gamma/\varepsilon_0$}}
     \put(155,39){{\footnotesize $\mathcal{T}_1$}}	
     \put(155,31){{\footnotesize $\mathcal{T}_2$}}	
     \put(155,23){{\footnotesize $\mathcal{T}_3$}}
\end{picture}
 \caption{The evolution of the field and electron parameters: (a) $E_x$, $p_y$ and $S_x$; (b) $E_y$, $p_x$ and $S_y$; (c) $S_z$ for  $g=g(\chi_e)$ (red solid), $g=2$ (blue dotted), and  $g=g(\chi_e)$ but
 without radiative polarization (magenta dashed),
  for an initially unpolarized electron beam. The electron spin dynamics is calculated numerically with Eq.(\ref{RPE}).
  At $t=0$  the peak of the laser pulse reaches the focal spot. The  field  and momentum components are normalized to their maximum. Radiation reaction for the electron momenta is neglected for simplicity  \cite{SM2}.
  (d) The values of the three terms in Eq.(\ref{dS}) are defined as $\mathcal{T}_1=\frac{1}{a}\{u^{2}\textrm{K}_{2/3}-u\textrm{K}_{1/3}[{\bm{S}}_{i}\cdot({\bm \beta} \times \bm s)]\}$, $\mathcal{T}_2=\frac{1}{a}\{u^{2}\textrm{IntK}_{1/3}-u\textrm{K}_{1/3}[{\bm{S}}_{i}\cdot({\bm \beta} \times \bm s)]\}$ and $\mathcal{T}_3=\frac{1}{a}u(1+u)\textrm{K}_{1/3}({\bm \beta} \times \bm s)$, with $a=(1+u)\mathrm{IntK_{1/3}}-(2+2u+u^2)\mathrm{K_{2/3}}+\bm{S}_i\cdot({\bm \beta} \times \bm s) u\mathrm{K_{1/3}}$ and $S_i=0.01$, $\bm{S}_i$ is parallel to $({\bm \beta} \times \bm s)$.} \label{fig3}
 \end{figure}

The reason for the electron beam polarization is analyzed in Fig.~\ref{fig3}. Without radiation reaction, the electrons typically move along a spiral trajectory in the CP laser pulse, with the transverse momentum perpendicular to the electric field [Figs.~\ref{fig3}(a,b)].
In each time step, the spin vector of an electron flips to the direction parallel or anti-parallel to the instantaneous spin quantization axis according to the quantum probabilities \cite{Liyfei2020}.
The evolution of polarization of an electron ensemble including both effects of spin-flip and spin precession
is described by the following equation \cite{Baier1972}:
\begin{eqnarray}\label{RPE}
\frac{d\bm{S}}{dt} &  =&{\bm S}\times {\bm F}-\frac{\alpha m}{\sqrt{3}\pi\gamma}\int_0^\infty\frac{u^{2}du}{\left(1+u\right)^{3}}\left(\mathrm{K_{2/3}}\bm{S}\nonumber\right.\\
&&\left.+\left(\mathrm{IntK_{1/3}}-\mathrm{K_{2/3}}\right)\left(\bm{S}\cdot\bm{\beta}\right)\bm{\beta}+\left(\bm{\beta}\times\bm{a}\right)\mathrm{K_{1/3}}\right),
\end{eqnarray}
where ${\rm IntK}_{\frac{1}{3}}\equiv \int_{u'}^{\infty} {\rm d}z {\rm K}_{\frac{1}{3}}(z)$,  ${\rm K}_{2/3}={\rm K}_{2/3}(u')$,  ${\rm K}_{1/3}={\rm K}_{1/3}(u')$,  ${\rm K}_n$ is the $n^{\rm th}$-order modified Bessel function of the second kind, $u'=2u/3\chi_e$,  $u=\omega_\gamma/(\varepsilon_0-\omega_\gamma)$,  $\varepsilon_0$ the electron energy, $\omega_\gamma$ the  photon energy, and $\bm{a}=\dot{\bm \beta}/|\dot{\bm \beta}|$.
As shown in Figs.~\ref{fig3} (a)-(c), the transverse polarization $\bm{S_\perp}$ oscillates synchronously with $\bm{E}$,  and $|S_z|$ builds up exponentially.
The phase matching between $S_{x,y}$ and $E_{x,y}$ follows from the domination of the last term in Eq. (\ref{RPE}) for
$\bm{S_\perp}$. In fact, $\mathrm{K_{2/3}}|\bm{S_\perp}|\ll \mathrm{K_{1/3}}$, and the spin precession role [first term in Eq.~(\ref{RPE})] is known to be minor for the transverse radiative polarization. Consequently,
$d\bm{S_\perp}/dt$ is parallel to $-\bm{\beta}\times\bm{a}$, and thus, $\bm{S_\perp}$ is parallel to $\bm{E}$.
We analyze the accumulation of the longitudinal polarization:
\begin{eqnarray}\label{paral}\nonumber
\frac{dS_\parallel}{dt}
&=&-\frac{e}{m}{\bm S_\perp} \cdot \left[\left(\frac{g}{2}-1\right) {{\bm { \beta}}} \times {\bf B} + \left(\frac{g \beta}{2}-\frac{1}{\beta}\right) {\bf E}\right]\nonumber\\
&& -\frac{\alpha m}{\sqrt{3}\pi\gamma} S_\parallel\int_0^\infty\frac{u^{2}du}{\left(1+u\right)^{3}}\mathrm{IntK_{1/3}},
\end{eqnarray}
with the use of Eq.~(11.171) in \cite{jackson1999classical}. For an initially unpolarized electrons beam, the longitudinal polarization arises due to the $\bm{S_\perp}$ term, connected to the spin procession, which can be approximated as $dS_\parallel/dt \approx -\frac{2e}{m}(\frac{g}{2}-1){\bm S_\perp} \cdot{\bf E}$, taking account of $S^i_\parallel=0$, $\beta\approx1$ and ${\bm \beta} \parallel-{\bm k}$.
As well known, the Dirac theory predicts precisely g=2, which leads to vanishing  longitudinal polarization  [Fig.\ref{fig3}(c)].
However,  the QED loop corrections induce an anomalous contribution to the electron's magnetic moment, $g\neq2$ [Eq.~(\ref{ggg})], which results in rotation of the transverse polarization to longitudinal direction in the case when $\bm{S}_\perp$ and $\bm{E}$ are phase matched. Note that, the phase matching  of $\bm{S}_\perp$ and $\bm{E}$ is a unique feature of electrons in a counter propagating circularly polarized laser field, and consequently such a configuration is essential for producing $P_\parallel$ \cite{SM}.
Therefore, the electron anomalous magnetic moment is the origin of helicity transfer. Our
estimation of ${\bm S_\parallel}$ via Eq.~(\ref{RPE}) yields $P_\parallel=3.26\%$ after the interaction, which is in accordance with the simulation result. Thus, even though the spin precession is trivial for the well-studied transverse polarization  in strong-field QED
 \cite{Liyf2019,Baier1972,Seipt2019}, it plays an essential role in generating longitudinal polarization. The correlation between anomalous magnetic moment and longitudinal polarization may consequently provide new potential of accurately measuring $g/2-1$.   For instance, the maximum and the changing rate of the averaged polarization degree exclusively depend on the anomalous magnetic moment, which would be sensitive measures of the anomaly $g/2-1$ \cite{rich1972current}.

Moreover, the Monte-Carlo simulation reveals a radial polarization feature of the electron beam with absolute value $P_\perp$ up to 65\% [Fig. \ref{fig1}(c)].  Even though the averaged transverse polarization is negligible due to the cancellation from opposite angles, which coincides with the prediction of modified BMT equations [Fig. \ref{fig3}(a,b)], it is possible to collect electrons in a certain angle to obtain a high transverse polarization,  as well as a high longitudinal polarization by applying a spin-rotating system \cite{Liyf2019}.
We can give a simple estimation of the radial polarization using the spin-flip transition probabilities, which determine
the spin change after a photon emission \cite{Liyf2019,Liyfei2020}:
\begin{widetext}
\begin{equation}\label{dS}
\Delta{\bm S^R} =\frac{\left[u^{2}\textrm{K}_{2/3}-u\textrm{K}_{1/3}\left[\bm{S}_{i}\cdot \left(\bm{\beta}\times\bm{a}\right)  \right]\right]\bm{S}_{\perp}
+\left[u^{2}\textrm{IntK}_{1/3}-u\textrm{K}_{1/3}\left[\bm{S}_{i}\cdot \left(\bm{\beta}\times\bm{a}\right)\right]\right]\bm{S}_{\parallel}
+u\left(1+u\right)\textrm{K}_{1/3}\left(\bm{\beta}\times\bm{a}\right)}{(1+u)\mathrm{IntK_{1/3}}-(2+2u+u^2)\mathrm{K_{2/3}}+\bm{S}_i\cdot({\bm \beta} \times \bm a) u\mathrm{K_{1/3}}}.
\end{equation}
\end{widetext}
The initial spin $\bm{S}_i$ is taken as the average polarization of the ensemble shown in Fig.\ref{fig3} (a)-(c), i.e.  $|{\bm P_\perp}| \sim 10^{-2}$ and $|{\bm P_\parallel}| \sim 10^{-2}$. 
In this case, the fist two terms of Eq. (\ref{dS}) are negligible compared with the last term since $|\bm{S}|\ll1$, and the remaining terms with modified Bessel functions are comparable, see Fig.\ref{fig3} (d).  Consequently, the change of transverse spin is estimated as
$\Delta{\bm S^R_\perp}\propto -{\bm \beta} \times \bm{a}$, anti-parallel to the momentum of the scattered electron  after one photon-emission \cite{SM}.
This feature is preserved during
multiple photon emissions, yielding the radial transverse electron polarization [Fig. \ref{fig1}(c)].

Since the transverse polarization $|\Delta\bm{S}^R_\perp|$ increases with the emitted photon energy \cite{SM},
the electrons that experience more energy loss obtain higher transverse polarization, which subsequently contributes to higher longitudinal polarization according to Eq.~(\ref{paral}). Meanwhile, as the deflection angle of electron $\theta_D\sim 1/\gamma$,
the transverse polarization degree increases with the decrease of $\theta$ [Fig.\ref{fig1}(d)]. Therefore, $P_\parallel$  is inversely proportional to $\varepsilon_e$ and $\theta$ [Figs.~\ref{fig1}(d),~\ref{fig2}].
The above analysis is not relevant
for scattered electrons with energy higher than 135 MeV, corresponding to those within angle of $\theta>167^\circ$. These electrons are polarized opposite to the laser helicity and the polarization degree increases with $\varepsilon_e$ and $\theta$ [Fig.~\ref{fig2}]. This counterintuitive  polarization feature highlights two different contributions of QED loop effects to the electron polarization in background fields. Firstly, the loop effects induce anomalous magnetic moment $g>2$, which yields the rotation of the radiative transverse polarization into the longitudinal direction. Secondly, the loop effects result in varying the electron spin even without photon emissions \cite{Meuren2011}. The latter effect is included in our Monte Carlo simulations by means of an additional no-photon-emission probability for the spin-flip \cite{CAIN}.  We estimate this effect via turning off artificially the photon-emission in the modified BMT equation \cite{SM}:
\begin{eqnarray}\label{Sno}\nonumber
\frac{d\bm{S}^{NR}}{dt}&=&\frac{e}{m}\left[\bm{S}\times\bm{F}\right]-\frac{\alpha m}{\sqrt{3}\pi\gamma}\int_{0}^{\infty}\frac{du}{\left(1+u\right)^{3}}\left\{\left[\bm{S}\cdot ({\bm \beta} \times \bm a)\nonumber\right.\right.\\
&&\left.\left.u\mathrm{K_{1/3}}\right]\bm{S}-({\bm \beta} \times \bm a) u\mathrm{K_{1/3}}\right\}.
\end{eqnarray}
Compared with Eq. (\ref{RPE}),
the dominant term $({\bm \beta} \times \bm a)$ has
opposite sign. Thus, if the transverse spin dynamics is governed by no-emission polarization,
the phase of ${\bm S}_\perp$ is opposite to  ${\bm E}$. Consequently, the energetic electrons obtain a negative longitudinal polarization [Fig.\ref{fig3}(c)] due to polarization rotation associated with anomalous magnetic moment, i.e.  $P_\parallel=-P_z\propto{\bf S_\perp}\cdot {\bf E}<0$, which is opposite to the low-energy electrons  in the remaining contribution to the spectrum, as shown in Fig. \ref{fig2}.

\begin{figure}
 	\includegraphics[width=1\linewidth]{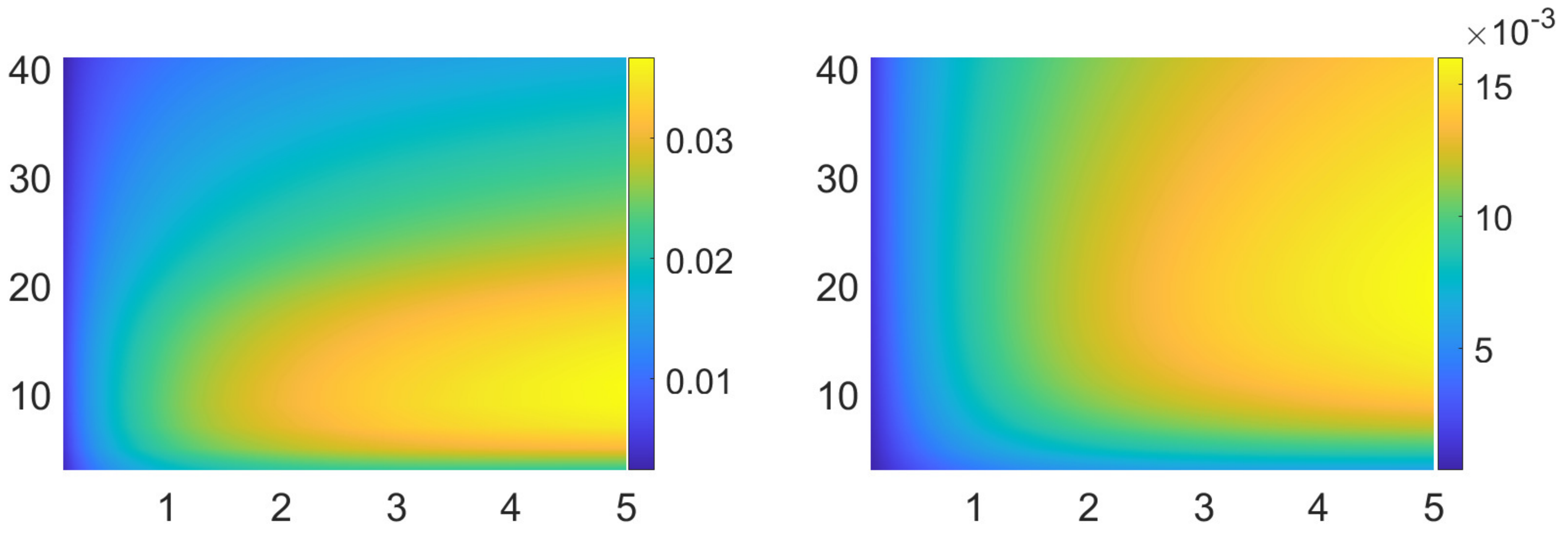}
 \begin{picture}(300,0)
  \put(13,75){\small {\color{white}(a)}}
 \put(-8,46){\rotatebox{90}{\small $\tau/T_0$}}	
 	 \put(52,3){{\small $\varepsilon_0$}}	
 	 \put(98,88){{\small $P_\parallel$}}	
 	
  \put(140,75){\small {\color{white}(b)}}
 \put(118,46){\rotatebox{90}{\small $\tau/T_0$}}	
 	 \put(181,3){{\small $\varepsilon_0$}}
 	 \put(220,88){{\small $P_\parallel$}}

\end{picture}
		 \caption{The dependence of the longitudinal polarization $P_\parallel$ on the laser duration $\tau$ and the initial electron energy $\varepsilon_0$ at the laser intensity of $a_0=100\sqrt{2}$ and $50 \sqrt{2}$, respectively.} \label{fig4}
 \end{figure}
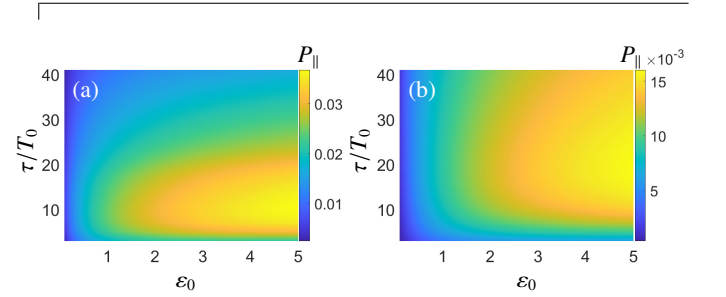

 The impacts of the laser and electron beam parameters on the longitudinal polarization are analyzed  in Fig.~\ref{fig4}.
With the  increase of $\varepsilon_0$ and $a_0$, the transverse polarization $P_\perp$  increases since radiative polarization is enhanced for a larger  radiation loss scaled by $\chi_e\approx 5\times10^{-6}a_0\gamma$ , and consequently,
the longitudinal polarization $P_\parallel$ grows. While for a certain $\varepsilon_0$ and $a_0$, $P_\parallel$ rises at first and then declines with respect to $\tau$. Without radiation reaction, one would expect a monotonously increase of $P_\parallel$ with the increase of interaction time from Eq.(\ref{RPE}). Unfortunately, radiation reaction breaks the phase correlation between ${\bm S}_\perp$ and ${\bm E}$, and disrupts the longitudinal polarization built at the preliminary stage of the interaction, resulting in the decrease of $P_\parallel$ for a long laser pulse \cite{SM}.

 Concluding, we have analyzed the role of  the anomalous magnetic moment and no-photon-emission spin dynamics,
  i.e. effects which both are consequences of QED radiative corrections,  for electron polarization  in the radiation dominated regime with multiple photon emissions.    We showed that exclusively due to these effects
    helicity transfer is possible from CP laser photons to electrons in ultrastrong  field regime $a_0\gg 1$,

which challenged the belief that circularly polarized laser beams cannot induce high longitudinal polarization.
This signature is robust with respect to the laser and electron parameters and  measurable with currently available experimental technology. It could serve for testing QED predictions on radiative corrections.


 {\it Acknowledgement:} The authors thank Y.-T. Li, W.-M. Wang and Q.-Z. Lv  for
helpful discussions. This work is supported by the National
Natural Science Foundation of China (Grants No. 12075187, and
No. 11804269), the Program for Professor of
Special Appointment (Eastern Scholar) at Shanghai
Institutions of Higher Learning, Shanghai Rising-Star
Program.

\bibliography{reference}

\begin{thebibliography}{73}%
\makeatletter
\providecommand \@ifxundefined [1]{%
 \@ifx{#1\undefined}
}%
\providecommand \@ifnum [1]{%
 \ifnum #1\expandafter \@firstoftwo
 \else \expandafter \@secondoftwo
 \fi
}%
\providecommand \@ifx [1]{%
 \ifx #1\expandafter \@firstoftwo
 \else \expandafter \@secondoftwo
 \fi
}%
\providecommand \natexlab [1]{#1}%
\providecommand \enquote  [1]{``#1''}%
\providecommand \bibnamefont  [1]{#1}%
\providecommand \bibfnamefont [1]{#1}%
\providecommand \citenamefont [1]{#1}%
\providecommand \href@noop [0]{\@secondoftwo}%
\providecommand \href [0]{\begingroup \@sanitize@url \@href}%
\providecommand \@href[1]{\@@startlink{#1}\@@href}%
\providecommand \@@href[1]{\endgroup#1\@@endlink}%
\providecommand \@sanitize@url [0]{\catcode `\\12\catcode `\$12\catcode
  `\&12\catcode `\#12\catcode `\^12\catcode `\_12\catcode `\%12\relax}%
\providecommand \@@startlink[1]{}%
\providecommand \@@endlink[0]{}%
\providecommand \url  [0]{\begingroup\@sanitize@url \@url }%
\providecommand \@url [1]{\endgroup\@href {#1}{\urlprefix }}%
\providecommand \urlprefix  [0]{URL }%
\providecommand \Eprint [0]{\href }%
\providecommand \doibase [0]{https://doi.org/}%
\providecommand \selectlanguage [0]{\@gobble}%
\providecommand \bibinfo  [0]{\@secondoftwo}%
\providecommand \bibfield  [0]{\@secondoftwo}%
\providecommand \translation [1]{[#1]}%
\providecommand \BibitemOpen [0]{}%
\providecommand \bibitemStop [0]{}%
\providecommand \bibitemNoStop [0]{.\EOS\space}%
\providecommand \EOS [0]{\spacefactor3000\relax}%
\providecommand \BibitemShut  [1]{\csname bibitem#1\endcsname}%
\let\auto@bib@innerbib\@empty
\bibitem [{\citenamefont {Yoon}\ \emph {et~al.}(2021)\citenamefont {Yoon},
  \citenamefont {Kim}, \citenamefont {Choi}, \citenamefont {Sung},
  \citenamefont {Lee}, \citenamefont {Lee},\ and\ \citenamefont
  {Nam}}]{Yoon2021}%
  \BibitemOpen
  \bibfield  {author} {\bibinfo {author} {\bibfnamefont {J.~W.}\ \bibnamefont
  {Yoon}}, \bibinfo {author} {\bibfnamefont {Y.~G.}\ \bibnamefont {Kim}},
  \bibinfo {author} {\bibfnamefont {I.~W.}\ \bibnamefont {Choi}}, \bibinfo
  {author} {\bibfnamefont {J.~H.}\ \bibnamefont {Sung}}, \bibinfo {author}
  {\bibfnamefont {H.~W.}\ \bibnamefont {Lee}}, \bibinfo {author} {\bibfnamefont
  {S.~K.}\ \bibnamefont {Lee}},\ and\ \bibinfo {author} {\bibfnamefont {C.~H.}\
  \bibnamefont {Nam}},\ }\bibfield  {title} {\bibinfo {title} {Realization of
  laser intensity over $10^{23}$ w/cm$^2$},\ }\href@noop {} {\bibfield
  {journal} {\bibinfo  {journal} {Optica}\ }\textbf {\bibinfo {volume} {8}},\
  \bibinfo {pages} {630} (\bibinfo {year} {2021})}\BibitemShut {NoStop}%
\bibitem [{\citenamefont {Schwinger}(1951)}]{schwinger1951gauge}%
  \BibitemOpen
  \bibfield  {author} {\bibinfo {author} {\bibfnamefont {J.}~\bibnamefont
  {Schwinger}},\ }\bibfield  {title} {\bibinfo {title} {On gauge invariance and
  vacuum polarization},\ }\href@noop {} {\bibfield  {journal} {\bibinfo
  {journal} {Physical Review}\ }\textbf {\bibinfo {volume} {82}},\ \bibinfo
  {pages} {664} (\bibinfo {year} {1951})}\BibitemShut {NoStop}%
\bibitem [{\citenamefont {Di~Piazza}\ \emph {et~al.}(2012)\citenamefont
  {Di~Piazza}, \citenamefont {M\"uller}, \citenamefont {Hatsagortsyan},\ and\
  \citenamefont {Keitel}}]{Piazza2012}%
  \BibitemOpen
  \bibfield  {author} {\bibinfo {author} {\bibfnamefont {A.}~\bibnamefont
  {Di~Piazza}}, \bibinfo {author} {\bibfnamefont {C.}~\bibnamefont {M\"uller}},
  \bibinfo {author} {\bibfnamefont {K.~Z.}\ \bibnamefont {Hatsagortsyan}},\
  and\ \bibinfo {author} {\bibfnamefont {C.~H.}\ \bibnamefont {Keitel}},\
  }\bibfield  {title} {\bibinfo {title} {Extremely high-intensity laser
  interactions with fundamental quantum systems},\ }\href
  {https://doi.org/10.1103/RevModPhys.84.1177} {\bibfield  {journal} {\bibinfo
  {journal} {Rev. Mod. Phys.}\ }\textbf {\bibinfo {volume} {84}},\ \bibinfo
  {pages} {1177} (\bibinfo {year} {2012})}\BibitemShut {NoStop}%
\bibitem [{\citenamefont {Jackson}(1999{\natexlab{a}})}]{jackson1999}%
  \BibitemOpen
  \bibfield  {author} {\bibinfo {author} {\bibfnamefont {J.~D.}\ \bibnamefont
  {Jackson}},\ }\href@noop {} {\emph {\bibinfo {title} {Classical
  electrodynamics}}}\ (\bibinfo  {publisher} {American Association of Physics
  Teachers},\ \bibinfo {year} {1999})\BibitemShut {NoStop}%
\bibitem [{\citenamefont {Abramowicz}\ \emph {et~al.}(2019)\citenamefont
  {Abramowicz}, \citenamefont {Altarelli}, \citenamefont {A{\ss}mann},
  \citenamefont {Behnke}, \citenamefont {Benhammou}, \citenamefont {Borysov},
  \citenamefont {Borysova}, \citenamefont {Brinkmann}, \citenamefont {Burkart},
  \citenamefont {B{\"u}{\ss}er} \emph {et~al.}}]{abramowicz2019letter}%
  \BibitemOpen
  \bibfield  {author} {\bibinfo {author} {\bibfnamefont {H.}~\bibnamefont
  {Abramowicz}}, \bibinfo {author} {\bibfnamefont {M.}~\bibnamefont
  {Altarelli}}, \bibinfo {author} {\bibfnamefont {R.}~\bibnamefont
  {A{\ss}mann}}, \bibinfo {author} {\bibfnamefont {T.}~\bibnamefont {Behnke}},
  \bibinfo {author} {\bibfnamefont {Y.}~\bibnamefont {Benhammou}}, \bibinfo
  {author} {\bibfnamefont {O.}~\bibnamefont {Borysov}}, \bibinfo {author}
  {\bibfnamefont {M.}~\bibnamefont {Borysova}}, \bibinfo {author}
  {\bibfnamefont {R.}~\bibnamefont {Brinkmann}}, \bibinfo {author}
  {\bibfnamefont {F.}~\bibnamefont {Burkart}}, \bibinfo {author} {\bibfnamefont
  {K.}~\bibnamefont {B{\"u}{\ss}er}}, \emph {et~al.},\ }\bibfield  {title}
  {\bibinfo {title} {Letter of intent for the luxe experiment},\ }\href@noop {}
  {\bibfield  {journal} {\bibinfo  {journal} {arXiv preprint arXiv:1909.00860}\
  } (\bibinfo {year} {2019})}\BibitemShut {NoStop}%
\bibitem [{unp()}]{unpublished}%
  \BibitemOpen
  \bibinfo {note} {The FACET-II SFQED Collaboration, Probing strong-field QED
  at FACET-II (SLAC E-320) (unpublished)}\BibitemShut {NoStop}%
\bibitem [{\citenamefont {Sokolov}\ and\ \citenamefont
  {Ternov}(1964)}]{Sokolov1964}%
  \BibitemOpen
  \bibfield  {author} {\bibinfo {author} {\bibfnamefont {A.~A.}\ \bibnamefont
  {Sokolov}}\ and\ \bibinfo {author} {\bibfnamefont {I.~M.}\ \bibnamefont
  {Ternov}},\ }\href@noop {} {\bibfield  {journal} {\bibinfo  {journal} {Sov.
  Phys. Dokl.}\ }\textbf {\bibinfo {volume} {8}},\ \bibinfo {pages} {1203}
  (\bibinfo {year} {1964})}\BibitemShut {NoStop}%
\bibitem [{\citenamefont {Baier}\ \emph {et~al.}(1971)\citenamefont {Baier},
  \citenamefont {Katkov},\ and\ \citenamefont
  {Strakhovenko}}]{baier1971radiative}%
  \BibitemOpen
  \bibfield  {author} {\bibinfo {author} {\bibfnamefont {V.}~\bibnamefont
  {Baier}}, \bibinfo {author} {\bibfnamefont {V.}~\bibnamefont {Katkov}},\ and\
  \bibinfo {author} {\bibfnamefont {V.}~\bibnamefont {Strakhovenko}},\
  }\bibfield  {title} {\bibinfo {title} {Radiative effects in an external
  electromagnetic field.},\ }in\ \href@noop {} {\emph {\bibinfo {booktitle}
  {Soviet Physics Doklady}}},\ Vol.~\bibinfo {volume} {16}\ (\bibinfo {year}
  {1971})\ p.\ \bibinfo {pages} {230}\BibitemShut {NoStop}%
\bibitem [{\citenamefont {Ba\ifmmode~\check{i}\else
  \v{i}\fi{}er}(1972)}]{Baier1972}%
  \BibitemOpen
  \bibfield  {author} {\bibinfo {author} {\bibfnamefont {V.~N.}\ \bibnamefont
  {Ba\ifmmode~\check{i}\else \v{i}\fi{}er}},\ }\bibfield  {title} {\bibinfo
  {title} {Radiative polarization of electrons in storage rings},\ }\href
  {http://stacks.iop.org/0038-5670/14/i=6/a=R02} {\bibfield  {journal}
  {\bibinfo  {journal} {Sov. Phys. Usp.}\ }\textbf {\bibinfo {volume} {14}},\
  \bibinfo {pages} {695} (\bibinfo {year} {1972})}\BibitemShut {NoStop}%
\bibitem [{\citenamefont {Derbenev}\ and\ \citenamefont
  {Kondratenko}(1973)}]{Derbenev_1973}%
  \BibitemOpen
  \bibfield  {author} {\bibinfo {author} {\bibfnamefont {Y.}~\bibnamefont
  {Derbenev}}\ and\ \bibinfo {author} {\bibfnamefont {A.~M.}\ \bibnamefont
  {Kondratenko}},\ }\bibfield  {title} {\bibinfo {title} {{Polarization
  kinematics of particles in storage rings}},\ }\href@noop {} {\bibfield
  {journal} {\bibinfo  {journal} {Zh. {\`{E}}ksper. Teoret. Fiz.}\ }\textbf
  {\bibinfo {volume} {64}},\ \bibinfo {pages} {1918} (\bibinfo {year}
  {1973})}\BibitemShut {NoStop}%
\bibitem [{\citenamefont {Baier}\ \emph {et~al.}(1998)\citenamefont {Baier},
  \citenamefont {Katkov},\ and\ \citenamefont {Strakhovenko}}]{Baier1998}%
  \BibitemOpen
  \bibfield  {author} {\bibinfo {author} {\bibfnamefont {V.~N.}\ \bibnamefont
  {Baier}}, \bibinfo {author} {\bibfnamefont {V.~M.}\ \bibnamefont {Katkov}},\
  and\ \bibinfo {author} {\bibfnamefont {V.~M.}\ \bibnamefont {Strakhovenko}},\
  }\href@noop {} {\emph {\bibinfo {title} {Electromagnetic Processes at High
  Energies in Oriented Single Crystals}}}\ (\bibinfo  {publisher} {World
  Scientific},\ \bibinfo {address} {Singapore},\ \bibinfo {year}
  {1998})\BibitemShut {NoStop}%
\bibitem [{\citenamefont {Li}\ \emph {et~al.}(2019)\citenamefont {Li},
  \citenamefont {Shaisultanov}, \citenamefont {Hatsagortsyan}, \citenamefont
  {Wan}, \citenamefont {Keitel},\ and\ \citenamefont {Li}}]{Liyf2019}%
  \BibitemOpen
  \bibfield  {author} {\bibinfo {author} {\bibfnamefont {Y.-F.}\ \bibnamefont
  {Li}}, \bibinfo {author} {\bibfnamefont {R.}~\bibnamefont {Shaisultanov}},
  \bibinfo {author} {\bibfnamefont {K.~Z.}\ \bibnamefont {Hatsagortsyan}},
  \bibinfo {author} {\bibfnamefont {F.}~\bibnamefont {Wan}}, \bibinfo {author}
  {\bibfnamefont {C.~H.}\ \bibnamefont {Keitel}},\ and\ \bibinfo {author}
  {\bibfnamefont {J.-X.}\ \bibnamefont {Li}},\ }\bibfield  {title} {\bibinfo
  {title} {Ultrarelativistic electron-beam polarization in single-shot
  interaction with an ultraintense laser pulse},\ }\href
  {https://doi.org/10.1103/PhysRevLett.122.154801} {\bibfield  {journal}
  {\bibinfo  {journal} {Phys. Rev. Lett.}\ }\textbf {\bibinfo {volume} {122}},\
  \bibinfo {pages} {154801} (\bibinfo {year} {2019})}\BibitemShut {NoStop}%
\bibitem [{\citenamefont {Seipt}\ \emph {et~al.}(2019)\citenamefont {Seipt},
  \citenamefont {Del~Sorbo}, \citenamefont {Ridgers},\ and\ \citenamefont
  {Thomas}}]{Seipt2019}%
  \BibitemOpen
  \bibfield  {author} {\bibinfo {author} {\bibfnamefont {D.}~\bibnamefont
  {Seipt}}, \bibinfo {author} {\bibfnamefont {D.}~\bibnamefont {Del~Sorbo}},
  \bibinfo {author} {\bibfnamefont {C.~P.}\ \bibnamefont {Ridgers}},\ and\
  \bibinfo {author} {\bibfnamefont {A.~G.~R.}\ \bibnamefont {Thomas}},\
  }\bibfield  {title} {\bibinfo {title} {Ultrafast polarization of an electron
  beam in an intense bichromatic laser field},\ }\href
  {https://doi.org/10.1103/PhysRevA.100.061402} {\bibfield  {journal} {\bibinfo
   {journal} {Phys. Rev. A}\ }\textbf {\bibinfo {volume} {100}},\ \bibinfo
  {pages} {061402} (\bibinfo {year} {2019})}\BibitemShut {NoStop}%
\bibitem [{\citenamefont {Li}\ \emph {et~al.}(2020{\natexlab{a}})\citenamefont
  {Li}, \citenamefont {Chen}, \citenamefont {Wang},\ and\ \citenamefont
  {Hu}}]{Liyfei2020}%
  \BibitemOpen
  \bibfield  {author} {\bibinfo {author} {\bibfnamefont {Y.-F.}\ \bibnamefont
  {Li}}, \bibinfo {author} {\bibfnamefont {Y.-Y.}\ \bibnamefont {Chen}},
  \bibinfo {author} {\bibfnamefont {W.-M.}\ \bibnamefont {Wang}},\ and\
  \bibinfo {author} {\bibfnamefont {H.-S.}\ \bibnamefont {Hu}},\ }\bibfield
  {title} {\bibinfo {title} {Production of highly polarized positron beams via
  helicity transfer from polarized electrons in a strong laser field},\ }\href
  {https://doi.org/10.1103/PhysRevLett.125.044802} {\bibfield  {journal}
  {\bibinfo  {journal} {Phys. Rev. Lett.}\ }\textbf {\bibinfo {volume} {125}},\
  \bibinfo {pages} {044802} (\bibinfo {year} {2020}{\natexlab{a}})}\BibitemShut
  {NoStop}%
\bibitem [{\citenamefont {Chen}\ \emph {et~al.}(2019)\citenamefont {Chen},
  \citenamefont {He}, \citenamefont {Shaisultanov}, \citenamefont
  {Hatsagortsyan},\ and\ \citenamefont {Keitel}}]{Chen2019}%
  \BibitemOpen
  \bibfield  {author} {\bibinfo {author} {\bibfnamefont {Y.-Y.}\ \bibnamefont
  {Chen}}, \bibinfo {author} {\bibfnamefont {P.-L.}\ \bibnamefont {He}},
  \bibinfo {author} {\bibfnamefont {R.}~\bibnamefont {Shaisultanov}}, \bibinfo
  {author} {\bibfnamefont {K.~Z.}\ \bibnamefont {Hatsagortsyan}},\ and\
  \bibinfo {author} {\bibfnamefont {C.~H.}\ \bibnamefont {Keitel}},\ }\bibfield
   {title} {\bibinfo {title} {Polarized positron beams via intense two-color
  laser pulses},\ }\href {https://doi.org/10.1103/PhysRevLett.123.174801}
  {\bibfield  {journal} {\bibinfo  {journal} {Phys. Rev. Lett.}\ }\textbf
  {\bibinfo {volume} {123}},\ \bibinfo {pages} {174801} (\bibinfo {year}
  {2019})}\BibitemShut {NoStop}%
\bibitem [{\citenamefont {Wan}\ \emph {et~al.}(2020)\citenamefont {Wan},
  \citenamefont {Shaisultanov}, \citenamefont {Li}, \citenamefont
  {Hatsagortsyan}, \citenamefont {Keitel},\ and\ \citenamefont {Li}}]{Wan2020}%
  \BibitemOpen
  \bibfield  {author} {\bibinfo {author} {\bibfnamefont {F.}~\bibnamefont
  {Wan}}, \bibinfo {author} {\bibfnamefont {R.}~\bibnamefont {Shaisultanov}},
  \bibinfo {author} {\bibfnamefont {Y.-F.}\ \bibnamefont {Li}}, \bibinfo
  {author} {\bibfnamefont {K.~Z.}\ \bibnamefont {Hatsagortsyan}}, \bibinfo
  {author} {\bibfnamefont {C.~H.}\ \bibnamefont {Keitel}},\ and\ \bibinfo
  {author} {\bibfnamefont {J.-X.}\ \bibnamefont {Li}},\ }\bibfield  {title}
  {\bibinfo {title} {Ultrarelativistic polarized positron jets via collision of
  electron and ultraintense laser beams},\ }\href
  {https://doi.org/https://doi.org/10.1016/j.physletb.2019.135120} {\bibfield
  {journal} {\bibinfo  {journal} {Physics Letters B}\ }\textbf {\bibinfo
  {volume} {800}},\ \bibinfo {pages} {135120} (\bibinfo {year}
  {2020})}\BibitemShut {NoStop}%
\bibitem [{\citenamefont {Li}\ \emph {et~al.}(2020{\natexlab{b}})\citenamefont
  {Li}, \citenamefont {Shaisultanov}, \citenamefont {Chen}, \citenamefont
  {Wan}, \citenamefont {Hatsagortsyan}, \citenamefont {Keitel},\ and\
  \citenamefont {Li}}]{Liyf2020}%
  \BibitemOpen
  \bibfield  {author} {\bibinfo {author} {\bibfnamefont {Y.-F.}\ \bibnamefont
  {Li}}, \bibinfo {author} {\bibfnamefont {R.}~\bibnamefont {Shaisultanov}},
  \bibinfo {author} {\bibfnamefont {Y.-Y.}\ \bibnamefont {Chen}}, \bibinfo
  {author} {\bibfnamefont {F.}~\bibnamefont {Wan}}, \bibinfo {author}
  {\bibfnamefont {K.~Z.}\ \bibnamefont {Hatsagortsyan}}, \bibinfo {author}
  {\bibfnamefont {C.~H.}\ \bibnamefont {Keitel}},\ and\ \bibinfo {author}
  {\bibfnamefont {J.-X.}\ \bibnamefont {Li}},\ }\bibfield  {title} {\bibinfo
  {title} {Polarized ultrashort brilliant multi-gev $\ensuremath{\gamma}$ rays
  via single-shot laser-electron interaction},\ }\href
  {https://doi.org/10.1103/PhysRevLett.124.014801} {\bibfield  {journal}
  {\bibinfo  {journal} {Phys. Rev. Lett.}\ }\textbf {\bibinfo {volume} {124}},\
  \bibinfo {pages} {014801} (\bibinfo {year} {2020}{\natexlab{b}})}\BibitemShut
  {NoStop}%
\bibitem [{\citenamefont {Tang}\ \emph {et~al.}(2020)\citenamefont {Tang},
  \citenamefont {King},\ and\ \citenamefont {Hu}}]{Tang2020}%
  \BibitemOpen
  \bibfield  {author} {\bibinfo {author} {\bibfnamefont {S.}~\bibnamefont
  {Tang}}, \bibinfo {author} {\bibfnamefont {B.}~\bibnamefont {King}},\ and\
  \bibinfo {author} {\bibfnamefont {H.}~\bibnamefont {Hu}},\ }\bibfield
  {title} {\bibinfo {title} {Highly polarised gamma photons from electron-laser
  collisions},\ }\href
  {https://doi.org/https://doi.org/10.1016/j.physletb.2020.135701} {\bibfield
  {journal} {\bibinfo  {journal} {Physics Letters B}\ }\textbf {\bibinfo
  {volume} {809}},\ \bibinfo {pages} {135701} (\bibinfo {year}
  {2020})}\BibitemShut {NoStop}%
\bibitem [{\citenamefont {Guo}\ \emph {et~al.}(2020)\citenamefont {Guo},
  \citenamefont {Wang}, \citenamefont {Shaisultanov}, \citenamefont {Wan},
  \citenamefont {Xu}, \citenamefont {Chen}, \citenamefont {Hatsagortsyan},\
  and\ \citenamefont {Li}}]{Guo2020}%
  \BibitemOpen
  \bibfield  {author} {\bibinfo {author} {\bibfnamefont {R.-T.}\ \bibnamefont
  {Guo}}, \bibinfo {author} {\bibfnamefont {Y.}~\bibnamefont {Wang}}, \bibinfo
  {author} {\bibfnamefont {R.}~\bibnamefont {Shaisultanov}}, \bibinfo {author}
  {\bibfnamefont {F.}~\bibnamefont {Wan}}, \bibinfo {author} {\bibfnamefont
  {Z.-F.}\ \bibnamefont {Xu}}, \bibinfo {author} {\bibfnamefont {Y.-Y.}\
  \bibnamefont {Chen}}, \bibinfo {author} {\bibfnamefont {K.~Z.}\ \bibnamefont
  {Hatsagortsyan}},\ and\ \bibinfo {author} {\bibfnamefont {J.-X.}\
  \bibnamefont {Li}},\ }\bibfield  {title} {\bibinfo {title} {Stochasticity in
  radiative polarization of ultrarelativistic electrons in an ultrastrong laser
  pulse},\ }\href {https://doi.org/10.1103/PhysRevResearch.2.033483} {\bibfield
   {journal} {\bibinfo  {journal} {Phys. Rev. Research}\ }\textbf {\bibinfo
  {volume} {2}},\ \bibinfo {pages} {033483} (\bibinfo {year}
  {2020})}\BibitemShut {NoStop}%
\bibitem [{\citenamefont {Gong}\ \emph {et~al.}(2021)\citenamefont {Gong},
  \citenamefont {Hatsagortsyan},\ and\ \citenamefont {Keitel}}]{Gong_2021}%
  \BibitemOpen
  \bibfield  {author} {\bibinfo {author} {\bibfnamefont {Z.}~\bibnamefont
  {Gong}}, \bibinfo {author} {\bibfnamefont {K.~Z.}\ \bibnamefont
  {Hatsagortsyan}},\ and\ \bibinfo {author} {\bibfnamefont {C.~H.}\
  \bibnamefont {Keitel}},\ }\bibfield  {title} {\bibinfo {title} {Retrieving
  transient magnetic fields of ultrarelativistic laser plasma via ejected
  electron polarization},\ }\href
  {https://doi.org/10.1103/PhysRevLett.127.165002} {\bibfield  {journal}
  {\bibinfo  {journal} {Phys. Rev. Lett.}\ }\textbf {\bibinfo {volume} {127}},\
  \bibinfo {pages} {165002} (\bibinfo {year} {2021})}\BibitemShut {NoStop}%
\bibitem [{\citenamefont {Moortgat-Pick}\ \emph {et~al.}(2008)\citenamefont
  {Moortgat-Pick}, \citenamefont {Abe}, \citenamefont {Alexander},
  \citenamefont {Ananthanarayan}, \citenamefont {Babich}, \citenamefont
  {Bharadwaj}, \citenamefont {Barber}, \citenamefont {Bartl}, \citenamefont
  {Brachmann}, \citenamefont {Chen}, \citenamefont {Clarke}, \citenamefont
  {Clendenin}, \citenamefont {Dainton}, \citenamefont {Desch}, \citenamefont
  {Diehl}, \citenamefont {Dobos}, \citenamefont {Dorland}, \citenamefont
  {Dreiner}, \citenamefont {Eberl}, \citenamefont {Ellis}, \citenamefont
  {Flöttmann}, \citenamefont {Fraas}, \citenamefont {Franco-Sollova},
  \citenamefont {Franke}, \citenamefont {Freitas}, \citenamefont {Goodson},
  \citenamefont {Gray}, \citenamefont {Han}, \citenamefont {Heinemeyer},
  \citenamefont {Hesselbach}, \citenamefont {Hirose}, \citenamefont
  {Hohenwarter-Sodek}, \citenamefont {Juste}, \citenamefont {Kalinowski},
  \citenamefont {Kernreiter}, \citenamefont {Kittel}, \citenamefont {Kraml},
  \citenamefont {Langenfeld}, \citenamefont {Majerotto}, \citenamefont
  {Martinez}, \citenamefont {Martyn}, \citenamefont {Mikhailichenko},
  \citenamefont {Milstene}, \citenamefont {Menges}, \citenamefont {Meyners},
  \citenamefont {Mönig}, \citenamefont {Moffeit}, \citenamefont {Moretti},
  \citenamefont {Nachtmann}, \citenamefont {Nagel}, \citenamefont {Nakanishi},
  \citenamefont {Nauenberg}, \citenamefont {Nowak}, \citenamefont {Omori},
  \citenamefont {Osland}, \citenamefont {Pankov}, \citenamefont {Paver},
  \citenamefont {Pitthan}, \citenamefont {Pöschl}, \citenamefont {Porod},
  \citenamefont {Proulx}, \citenamefont {Richardson}, \citenamefont {Riemann},
  \citenamefont {Rindani}, \citenamefont {Rizzo}, \citenamefont {Schälicke},
  \citenamefont {Schüler}, \citenamefont {Schwanenberger}, \citenamefont
  {Scott}, \citenamefont {Sheppard}, \citenamefont {Singh}, \citenamefont
  {Sopczak}, \citenamefont {Spiesberger}, \citenamefont {Stahl}, \citenamefont
  {Steiner}, \citenamefont {Wagner}, \citenamefont {Weber}, \citenamefont
  {Weiglein}, \citenamefont {Wilson}, \citenamefont {Woods}, \citenamefont
  {Zerwas}, \citenamefont {Zhang},\ and\ \citenamefont {Zomer}}]{Moortgat2008}%
  \BibitemOpen
  \bibfield  {author} {\bibinfo {author} {\bibfnamefont {G.}~\bibnamefont
  {Moortgat-Pick}}, \bibinfo {author} {\bibfnamefont {T.}~\bibnamefont {Abe}},
  \bibinfo {author} {\bibfnamefont {G.}~\bibnamefont {Alexander}}, \bibinfo
  {author} {\bibfnamefont {B.}~\bibnamefont {Ananthanarayan}}, \bibinfo
  {author} {\bibfnamefont {A.}~\bibnamefont {Babich}}, \bibinfo {author}
  {\bibfnamefont {V.}~\bibnamefont {Bharadwaj}}, \bibinfo {author}
  {\bibfnamefont {D.}~\bibnamefont {Barber}}, \bibinfo {author} {\bibfnamefont
  {A.}~\bibnamefont {Bartl}}, \bibinfo {author} {\bibfnamefont
  {A.}~\bibnamefont {Brachmann}}, \bibinfo {author} {\bibfnamefont
  {S.}~\bibnamefont {Chen}}, \bibinfo {author} {\bibfnamefont {J.}~\bibnamefont
  {Clarke}}, \bibinfo {author} {\bibfnamefont {J.}~\bibnamefont {Clendenin}},
  \bibinfo {author} {\bibfnamefont {J.}~\bibnamefont {Dainton}}, \bibinfo
  {author} {\bibfnamefont {K.}~\bibnamefont {Desch}}, \bibinfo {author}
  {\bibfnamefont {M.}~\bibnamefont {Diehl}}, \bibinfo {author} {\bibfnamefont
  {B.}~\bibnamefont {Dobos}}, \bibinfo {author} {\bibfnamefont
  {T.}~\bibnamefont {Dorland}}, \bibinfo {author} {\bibfnamefont
  {H.}~\bibnamefont {Dreiner}}, \bibinfo {author} {\bibfnamefont
  {H.}~\bibnamefont {Eberl}}, \bibinfo {author} {\bibfnamefont
  {J.}~\bibnamefont {Ellis}}, \bibinfo {author} {\bibfnamefont
  {K.}~\bibnamefont {Flöttmann}}, \bibinfo {author} {\bibfnamefont
  {H.}~\bibnamefont {Fraas}}, \bibinfo {author} {\bibfnamefont
  {F.}~\bibnamefont {Franco-Sollova}}, \bibinfo {author} {\bibfnamefont
  {F.}~\bibnamefont {Franke}}, \bibinfo {author} {\bibfnamefont
  {A.}~\bibnamefont {Freitas}}, \bibinfo {author} {\bibfnamefont
  {J.}~\bibnamefont {Goodson}}, \bibinfo {author} {\bibfnamefont
  {J.}~\bibnamefont {Gray}}, \bibinfo {author} {\bibfnamefont {A.}~\bibnamefont
  {Han}}, \bibinfo {author} {\bibfnamefont {S.}~\bibnamefont {Heinemeyer}},
  \bibinfo {author} {\bibfnamefont {S.}~\bibnamefont {Hesselbach}}, \bibinfo
  {author} {\bibfnamefont {T.}~\bibnamefont {Hirose}}, \bibinfo {author}
  {\bibfnamefont {K.}~\bibnamefont {Hohenwarter-Sodek}}, \bibinfo {author}
  {\bibfnamefont {A.}~\bibnamefont {Juste}}, \bibinfo {author} {\bibfnamefont
  {J.}~\bibnamefont {Kalinowski}}, \bibinfo {author} {\bibfnamefont
  {T.}~\bibnamefont {Kernreiter}}, \bibinfo {author} {\bibfnamefont
  {O.}~\bibnamefont {Kittel}}, \bibinfo {author} {\bibfnamefont
  {S.}~\bibnamefont {Kraml}}, \bibinfo {author} {\bibfnamefont
  {U.}~\bibnamefont {Langenfeld}}, \bibinfo {author} {\bibfnamefont
  {W.}~\bibnamefont {Majerotto}}, \bibinfo {author} {\bibfnamefont
  {A.}~\bibnamefont {Martinez}}, \bibinfo {author} {\bibfnamefont {H.-U.}\
  \bibnamefont {Martyn}}, \bibinfo {author} {\bibfnamefont {A.}~\bibnamefont
  {Mikhailichenko}}, \bibinfo {author} {\bibfnamefont {C.}~\bibnamefont
  {Milstene}}, \bibinfo {author} {\bibfnamefont {W.}~\bibnamefont {Menges}},
  \bibinfo {author} {\bibfnamefont {N.}~\bibnamefont {Meyners}}, \bibinfo
  {author} {\bibfnamefont {K.}~\bibnamefont {Mönig}}, \bibinfo {author}
  {\bibfnamefont {K.}~\bibnamefont {Moffeit}}, \bibinfo {author} {\bibfnamefont
  {S.}~\bibnamefont {Moretti}}, \bibinfo {author} {\bibfnamefont
  {O.}~\bibnamefont {Nachtmann}}, \bibinfo {author} {\bibfnamefont
  {F.}~\bibnamefont {Nagel}}, \bibinfo {author} {\bibfnamefont
  {T.}~\bibnamefont {Nakanishi}}, \bibinfo {author} {\bibfnamefont
  {U.}~\bibnamefont {Nauenberg}}, \bibinfo {author} {\bibfnamefont
  {H.}~\bibnamefont {Nowak}}, \bibinfo {author} {\bibfnamefont
  {T.}~\bibnamefont {Omori}}, \bibinfo {author} {\bibfnamefont
  {P.}~\bibnamefont {Osland}}, \bibinfo {author} {\bibfnamefont
  {A.}~\bibnamefont {Pankov}}, \bibinfo {author} {\bibfnamefont
  {N.}~\bibnamefont {Paver}}, \bibinfo {author} {\bibfnamefont
  {R.}~\bibnamefont {Pitthan}}, \bibinfo {author} {\bibfnamefont
  {R.}~\bibnamefont {Pöschl}}, \bibinfo {author} {\bibfnamefont
  {W.}~\bibnamefont {Porod}}, \bibinfo {author} {\bibfnamefont
  {J.}~\bibnamefont {Proulx}}, \bibinfo {author} {\bibfnamefont
  {P.}~\bibnamefont {Richardson}}, \bibinfo {author} {\bibfnamefont
  {S.}~\bibnamefont {Riemann}}, \bibinfo {author} {\bibfnamefont
  {S.}~\bibnamefont {Rindani}}, \bibinfo {author} {\bibfnamefont
  {T.}~\bibnamefont {Rizzo}}, \bibinfo {author} {\bibfnamefont
  {A.}~\bibnamefont {Schälicke}}, \bibinfo {author} {\bibfnamefont
  {P.}~\bibnamefont {Schüler}}, \bibinfo {author} {\bibfnamefont
  {C.}~\bibnamefont {Schwanenberger}}, \bibinfo {author} {\bibfnamefont
  {D.}~\bibnamefont {Scott}}, \bibinfo {author} {\bibfnamefont
  {J.}~\bibnamefont {Sheppard}}, \bibinfo {author} {\bibfnamefont
  {R.}~\bibnamefont {Singh}}, \bibinfo {author} {\bibfnamefont
  {A.}~\bibnamefont {Sopczak}}, \bibinfo {author} {\bibfnamefont
  {H.}~\bibnamefont {Spiesberger}}, \bibinfo {author} {\bibfnamefont
  {A.}~\bibnamefont {Stahl}}, \bibinfo {author} {\bibfnamefont
  {H.}~\bibnamefont {Steiner}}, \bibinfo {author} {\bibfnamefont
  {A.}~\bibnamefont {Wagner}}, \bibinfo {author} {\bibfnamefont
  {A.}~\bibnamefont {Weber}}, \bibinfo {author} {\bibfnamefont
  {G.}~\bibnamefont {Weiglein}}, \bibinfo {author} {\bibfnamefont
  {G.}~\bibnamefont {Wilson}}, \bibinfo {author} {\bibfnamefont
  {M.}~\bibnamefont {Woods}}, \bibinfo {author} {\bibfnamefont
  {P.}~\bibnamefont {Zerwas}}, \bibinfo {author} {\bibfnamefont
  {J.}~\bibnamefont {Zhang}},\ and\ \bibinfo {author} {\bibfnamefont
  {F.}~\bibnamefont {Zomer}},\ }\bibfield  {title} {\bibinfo {title} {Polarized
  positrons and electrons at the linear collider},\ }\href@noop {} {\bibfield
  {journal} {\bibinfo  {journal} {Phys. Rep.}\ }\textbf {\bibinfo {volume}
  {460}},\ \bibinfo {pages} {131 } (\bibinfo {year} {2008})}\BibitemShut
  {NoStop}%
\bibitem [{\citenamefont {Subashiev}\ \emph {et~al.}(1998)\citenamefont
  {Subashiev}, \citenamefont {Yashin}, \citenamefont {Clendenin},\ and\
  \citenamefont {A.Mamaev}}]{Subashiev1998}%
  \BibitemOpen
  \bibfield  {author} {\bibinfo {author} {\bibfnamefont {A.~V.}\ \bibnamefont
  {Subashiev}}, \bibinfo {author} {\bibfnamefont {Y.~P.}\ \bibnamefont
  {Yashin}}, \bibinfo {author} {\bibfnamefont {J.~E.}\ \bibnamefont
  {Clendenin}},\ and\ \bibinfo {author} {\bibfnamefont {Y.}~\bibnamefont
  {A.Mamaev}},\ }\bibfield  {title} {\bibinfo {title} {Spin polarized
  electrons: Generation and applications},\ }\href@noop {} {\bibfield
  {journal} {\bibinfo  {journal} {Phys. Low Dimens. Struct.}\ }\textbf
  {\bibinfo {volume} {1}} (\bibinfo {year} {1998})},\ \Eprint
  {https://arxiv.org/abs/[SLAC PUB 8035 (1998)]} {[SLAC PUB 8035 (1998)]}
  \BibitemShut {NoStop}%
\bibitem [{\citenamefont {Elouadrhiri}\ \emph {et~al.}(2009)\citenamefont
  {Elouadrhiri}, \citenamefont {Forest}, \citenamefont {Grames}, \citenamefont
  {Melnitchouk},\ and\ \citenamefont {Voutier}}]{Elouadrhiri2009}%
  \BibitemOpen
  \bibfield  {author} {\bibinfo {author} {\bibfnamefont {L.}~\bibnamefont
  {Elouadrhiri}}, \bibinfo {author} {\bibfnamefont {T.~A.}\ \bibnamefont
  {Forest}}, \bibinfo {author} {\bibfnamefont {J.}~\bibnamefont {Grames}},
  \bibinfo {author} {\bibfnamefont {W.}~\bibnamefont {Melnitchouk}},\ and\
  \bibinfo {author} {\bibfnamefont {E.}~\bibnamefont {Voutier}},\ }\bibfield
  {title} {\bibinfo {title} {Proceedings of the international workshop on
  positrons at jefferson lab},\ }\href@noop {} {\bibfield  {journal} {\bibinfo
  {journal} {AIP Conf. Proc.}\ }\textbf {\bibinfo {volume} {1160}} (\bibinfo
  {year} {2009})}\BibitemShut {NoStop}%
\bibitem [{\citenamefont {Rich}\ \emph {et~al.}(1987)\citenamefont {Rich},
  \citenamefont {House}, \citenamefont {Gidley},\ and\ \citenamefont
  {Conti}}]{Rich1987}%
  \BibitemOpen
  \bibfield  {author} {\bibinfo {author} {\bibfnamefont {A.}~\bibnamefont
  {Rich}}, \bibinfo {author} {\bibfnamefont {J.~V.}\ \bibnamefont {House}},
  \bibinfo {author} {\bibfnamefont {D.~W.}\ \bibnamefont {Gidley}},\ and\
  \bibinfo {author} {\bibfnamefont {R.~S.}\ \bibnamefont {Conti}},\ }\bibfield
  {title} {\bibinfo {title} {Spin-polarized low-energy positron beams and their
  applications},\ }\href@noop {} {\bibfield  {journal} {\bibinfo  {journal}
  {Appl. Phys. A}\ }\textbf {\bibinfo {volume} {43}},\ \bibinfo {pages} {275}
  (\bibinfo {year} {1987})}\BibitemShut {NoStop}%
\bibitem [{\citenamefont {Gidley}\ \emph {et~al.}(1982)\citenamefont {Gidley},
  \citenamefont {K\"oymen},\ and\ \citenamefont {Capehart}}]{Gidley1982}%
  \BibitemOpen
  \bibfield  {author} {\bibinfo {author} {\bibfnamefont {D.~W.}\ \bibnamefont
  {Gidley}}, \bibinfo {author} {\bibfnamefont {A.~R.}\ \bibnamefont
  {K\"oymen}},\ and\ \bibinfo {author} {\bibfnamefont {T.~W.}\ \bibnamefont
  {Capehart}},\ }\bibfield  {title} {\bibinfo {title} {Polarized low-energy
  positrons: A new probe of surface magnetism},\ }\href
  {https://doi.org/10.1103/PhysRevLett.49.1779} {\bibfield  {journal} {\bibinfo
   {journal} {Phys. Rev. Lett.}\ }\textbf {\bibinfo {volume} {49}},\ \bibinfo
  {pages} {1779} (\bibinfo {year} {1982})}\BibitemShut {NoStop}%
\bibitem [{\citenamefont {Ivanov}\ \emph {et~al.}(2004)\citenamefont {Ivanov},
  \citenamefont {Kotkin},\ and\ \citenamefont {Serbo}}]{Ivanov_2004}%
  \BibitemOpen
  \bibfield  {author} {\bibinfo {author} {\bibfnamefont {D.~Y.}\ \bibnamefont
  {Ivanov}}, \bibinfo {author} {\bibfnamefont {G.~L.}\ \bibnamefont {Kotkin}},\
  and\ \bibinfo {author} {\bibfnamefont {V.~G.}\ \bibnamefont {Serbo}},\
  }\bibfield  {title} {\bibinfo {title} {{Complete description of polarization
  effects in emission of a photon by an electron in the field of a strong laser
  wave}},\ }\href@noop {} {\bibfield  {journal} {\bibinfo  {journal} {Eur.
  Phys. J. C}\ }\textbf {\bibinfo {volume} {36}},\ \bibinfo {pages} {127}
  (\bibinfo {year} {2004})}\BibitemShut {NoStop}%
\bibitem [{\citenamefont {Seipt}\ and\ \citenamefont {King}(2020)}]{Seipt2020}%
  \BibitemOpen
  \bibfield  {author} {\bibinfo {author} {\bibfnamefont {D.}~\bibnamefont
  {Seipt}}\ and\ \bibinfo {author} {\bibfnamefont {B.}~\bibnamefont {King}},\
  }\bibfield  {title} {\bibinfo {title} {Spin- and polarization-dependent
  locally-constant-field-approximation rates for nonlinear compton and
  breit-wheeler processes},\ }\href
  {https://doi.org/10.1103/PhysRevA.102.052805} {\bibfield  {journal} {\bibinfo
   {journal} {Phys. Rev. A}\ }\textbf {\bibinfo {volume} {102}},\ \bibinfo
  {pages} {052805} (\bibinfo {year} {2020})}\BibitemShut {NoStop}%
\bibitem [{\citenamefont {Wistisen}\ and\ \citenamefont
  {Di~Piazza}(2019)}]{wistisen2019numerical}%
  \BibitemOpen
  \bibfield  {author} {\bibinfo {author} {\bibfnamefont {T.}~\bibnamefont
  {Wistisen}}\ and\ \bibinfo {author} {\bibfnamefont {A.}~\bibnamefont
  {Di~Piazza}},\ }\bibfield  {title} {\bibinfo {title} {Numerical approach to
  the semiclassical method of radiation emission for arbitrary electron spin
  and photon polarization},\ }\href@noop {} {\bibfield  {journal} {\bibinfo
  {journal} {Physical Review D}\ }\textbf {\bibinfo {volume} {100}},\ \bibinfo
  {pages} {116001} (\bibinfo {year} {2019})}\BibitemShut {NoStop}%
\bibitem [{\citenamefont {King}\ and\ \citenamefont
  {Tang}(2020)}]{king2020nonlinear}%
  \BibitemOpen
  \bibfield  {author} {\bibinfo {author} {\bibfnamefont {B.}~\bibnamefont
  {King}}\ and\ \bibinfo {author} {\bibfnamefont {S.}~\bibnamefont {Tang}},\
  }\bibfield  {title} {\bibinfo {title} {Nonlinear compton scattering of
  polarized photons in plane-wave backgrounds},\ }\href@noop {} {\bibfield
  {journal} {\bibinfo  {journal} {Physical Review A}\ }\textbf {\bibinfo
  {volume} {102}},\ \bibinfo {pages} {022809} (\bibinfo {year}
  {2020})}\BibitemShut {NoStop}%
\bibitem [{\citenamefont {Baier}\ \emph {et~al.}(1975)\citenamefont {Baier},
  \citenamefont {Katkov}, \citenamefont {Milshtein},\ and\ \citenamefont
  {Strakhovenko}}]{Baier_1975}%
  \BibitemOpen
  \bibfield  {author} {\bibinfo {author} {\bibfnamefont {V.~N.}\ \bibnamefont
  {Baier}}, \bibinfo {author} {\bibfnamefont {V.~M.}\ \bibnamefont {Katkov}},
  \bibinfo {author} {\bibfnamefont {A.~I.}\ \bibnamefont {Milshtein}},\ and\
  \bibinfo {author} {\bibnamefont {Strakhovenko}},\ }\bibfield  {title}
  {\bibinfo {title} {{The theory of quantum processes in the field of a strong
  electromagnetic wave}},\ }\href@noop {} {\bibfield  {journal} {\bibinfo
  {journal} {Zh. Eksp. Teor. Fiz}\ }\textbf {\bibinfo {volume} {69}},\ \bibinfo
  {pages} {783} (\bibinfo {year} {1975})}\BibitemShut {NoStop}%
\bibitem [{\citenamefont {Meuren}\ and\ \citenamefont
  {Di~Piazza}(2011)}]{Meuren2011}%
  \BibitemOpen
  \bibfield  {author} {\bibinfo {author} {\bibfnamefont {S.}~\bibnamefont
  {Meuren}}\ and\ \bibinfo {author} {\bibfnamefont {A.}~\bibnamefont
  {Di~Piazza}},\ }\bibfield  {title} {\bibinfo {title} {Quantum electron
  self-interaction in a strong laser field},\ }\href
  {https://doi.org/10.1103/PhysRevLett.107.260401} {\bibfield  {journal}
  {\bibinfo  {journal} {Phys. Rev. Lett.}\ }\textbf {\bibinfo {volume} {107}},\
  \bibinfo {pages} {260401} (\bibinfo {year} {2011})}\BibitemShut {NoStop}%
\bibitem [{\citenamefont {Podszus}\ and\ \citenamefont
  {Di~Piazza}(2021)}]{podszus2021first}%
  \BibitemOpen
  \bibfield  {author} {\bibinfo {author} {\bibfnamefont {T.}~\bibnamefont
  {Podszus}}\ and\ \bibinfo {author} {\bibfnamefont {A.}~\bibnamefont
  {Di~Piazza}},\ }\bibfield  {title} {\bibinfo {title} {First-order
  strong-field qed processes including the damping of particle states},\
  }\href@noop {} {\bibfield  {journal} {\bibinfo  {journal} {Physical Review
  D}\ }\textbf {\bibinfo {volume} {104}},\ \bibinfo {pages} {016014} (\bibinfo
  {year} {2021})}\BibitemShut {NoStop}%
\bibitem [{\citenamefont {Ilderton}\ \emph {et~al.}(2020)\citenamefont
  {Ilderton}, \citenamefont {King},\ and\ \citenamefont {Tang}}]{Ilderton2020}%
  \BibitemOpen
  \bibfield  {author} {\bibinfo {author} {\bibfnamefont {A.}~\bibnamefont
  {Ilderton}}, \bibinfo {author} {\bibfnamefont {B.}~\bibnamefont {King}},\
  and\ \bibinfo {author} {\bibfnamefont {S.}~\bibnamefont {Tang}},\ }\bibfield
  {title} {\bibinfo {title} {Loop spin effects in intense background fields},\
  }\href {https://doi.org/10.1103/PhysRevD.102.076013} {\bibfield  {journal}
  {\bibinfo  {journal} {Phys. Rev. D}\ }\textbf {\bibinfo {volume} {102}},\
  \bibinfo {pages} {076013} (\bibinfo {year} {2020})}\BibitemShut {NoStop}%
\bibitem [{\citenamefont {Torgrimsson}(2021{\natexlab{a}})}]{Torgrimsson_2021}%
  \BibitemOpen
  \bibfield  {author} {\bibinfo {author} {\bibfnamefont {G.}~\bibnamefont
  {Torgrimsson}},\ }\bibfield  {title} {\bibinfo {title} {Loops and
  polarization in strong-field {QED}},\ }\href
  {https://doi.org/10.1088/1367-2630/abf274} {\bibfield  {journal} {\bibinfo
  {journal} {New J. Phys.}\ }\textbf {\bibinfo {volume} {23}},\ \bibinfo
  {pages} {065001} (\bibinfo {year} {2021}{\natexlab{a}})}\BibitemShut
  {NoStop}%
\bibitem [{\citenamefont
  {Torgrimsson}(2021{\natexlab{b}})}]{torgrimsson2021resummation}%
  \BibitemOpen
  \bibfield  {author} {\bibinfo {author} {\bibfnamefont {G.}~\bibnamefont
  {Torgrimsson}},\ }\bibfield  {title} {\bibinfo {title} {Resummation of
  quantum radiation reaction and induced polarization},\ }\href@noop {}
  {\bibfield  {journal} {\bibinfo  {journal} {Physical Review D}\ }\textbf
  {\bibinfo {volume} {104}},\ \bibinfo {pages} {056016} (\bibinfo {year}
  {2021}{\natexlab{b}})}\BibitemShut {NoStop}%
\bibitem [{\citenamefont {Seipt}\ \emph {et~al.}(2021)\citenamefont {Seipt},
  \citenamefont {Ridgers}, \citenamefont {Sorbo},\ and\ \citenamefont
  {Thomas}}]{Seipt2021}%
  \BibitemOpen
  \bibfield  {author} {\bibinfo {author} {\bibfnamefont {D.}~\bibnamefont
  {Seipt}}, \bibinfo {author} {\bibfnamefont {C.~P.}\ \bibnamefont {Ridgers}},
  \bibinfo {author} {\bibfnamefont {D.~D.}\ \bibnamefont {Sorbo}},\ and\
  \bibinfo {author} {\bibfnamefont {A.~G.~R.}\ \bibnamefont {Thomas}},\
  }\bibfield  {title} {\bibinfo {title} {Polarized {QED} cascades},\ }\href
  {https://doi.org/10.1088/1367-2630/abf584} {\bibfield  {journal} {\bibinfo
  {journal} {New Journal of Physics}\ }\textbf {\bibinfo {volume} {23}},\
  \bibinfo {pages} {053025} (\bibinfo {year} {2021})}\BibitemShut {NoStop}%
\bibitem [{\citenamefont {Carlini}\ \emph {et~al.}(2012)\citenamefont
  {Carlini}, \citenamefont {Finn}, \citenamefont {Kowalski}, \citenamefont
  {Page}, \citenamefont {Armstrong}, \citenamefont {Asaturyan}, \citenamefont
  {Averett}, \citenamefont {Benesch}, \citenamefont {Birchall}, \citenamefont
  {Bosted} \emph {et~al.}}]{carlini2012qweak}%
  \BibitemOpen
  \bibfield  {author} {\bibinfo {author} {\bibfnamefont {R.~D.}\ \bibnamefont
  {Carlini}}, \bibinfo {author} {\bibfnamefont {J.}~\bibnamefont {Finn}},
  \bibinfo {author} {\bibfnamefont {S.}~\bibnamefont {Kowalski}}, \bibinfo
  {author} {\bibfnamefont {S.}~\bibnamefont {Page}}, \bibinfo {author}
  {\bibfnamefont {D.}~\bibnamefont {Armstrong}}, \bibinfo {author}
  {\bibfnamefont {A.}~\bibnamefont {Asaturyan}}, \bibinfo {author}
  {\bibfnamefont {T.}~\bibnamefont {Averett}}, \bibinfo {author} {\bibfnamefont
  {J.}~\bibnamefont {Benesch}}, \bibinfo {author} {\bibfnamefont
  {J.}~\bibnamefont {Birchall}}, \bibinfo {author} {\bibfnamefont
  {P.}~\bibnamefont {Bosted}}, \emph {et~al.},\ }\bibfield  {title} {\bibinfo
  {title} {The qweak experiment: A search for new physics at the tev scale via
  a measurement of the proton's weak charge},\ }\href@noop {} {\bibfield
  {journal} {\bibinfo  {journal} {arXiv preprint arXiv:1202.1255}\ } (\bibinfo
  {year} {2012})}\BibitemShut {NoStop}%
\bibitem [{\citenamefont {Kowalski}(2020)}]{kowalski2020parity}%
  \BibitemOpen
  \bibfield  {author} {\bibinfo {author} {\bibfnamefont {S.}~\bibnamefont
  {Kowalski}},\ }\bibfield  {title} {\bibinfo {title} {Parity-violating
  electron scattering},\ }\href@noop {} {\bibfield  {journal} {\bibinfo
  {journal} {HNPS Advances in Nuclear Physics}\ }\textbf {\bibinfo {volume}
  {15}},\ \bibinfo {pages} {2} (\bibinfo {year} {2020})}\BibitemShut {NoStop}%
\bibitem [{\citenamefont {Pierce}\ and\ \citenamefont
  {Meier}(1976)}]{Pierce_1976}%
  \BibitemOpen
  \bibfield  {author} {\bibinfo {author} {\bibfnamefont {D.~T.}\ \bibnamefont
  {Pierce}}\ and\ \bibinfo {author} {\bibfnamefont {F.}~\bibnamefont {Meier}},\
  }\bibfield  {title} {\bibinfo {title} {Photoemission of spin-polarized
  electrons from gaas},\ }\href {https://doi.org/10.1103/PhysRevB.13.5484}
  {\bibfield  {journal} {\bibinfo  {journal} {Phys. Rev. B}\ }\textbf {\bibinfo
  {volume} {13}},\ \bibinfo {pages} {5484} (\bibinfo {year}
  {1976})}\BibitemShut {NoStop}%
\bibitem [{\citenamefont {Swartz}(1988)}]{Swartz_1988}%
  \BibitemOpen
  \bibfield  {author} {\bibinfo {author} {\bibfnamefont {M.}~\bibnamefont
  {Swartz}},\ }\bibfield  {title} {\bibinfo {title} {Physics with polarized
  beams},\ }\href@noop {} {\bibfield  {journal} {\bibinfo  {journal} {Report
  No. SLAC-PUB-4656}\ } (\bibinfo {year} {1988})}\BibitemShut {NoStop}%
\bibitem [{\citenamefont {Ohgaki}\ \emph {et~al.}(1996)\citenamefont {Ohgaki},
  \citenamefont {Noguchi}, \citenamefont {Sugiyama}, \citenamefont {Mikado},
  \citenamefont {Chiwaki}, \citenamefont {Yamada}, \citenamefont {Suzuki},
  \citenamefont {Sei}, \citenamefont {Ohdaira},\ and\ \citenamefont
  {Yamazaki}}]{Ohgaki_1996}%
  \BibitemOpen
  \bibfield  {author} {\bibinfo {author} {\bibfnamefont {H.}~\bibnamefont
  {Ohgaki}}, \bibinfo {author} {\bibfnamefont {T.}~\bibnamefont {Noguchi}},
  \bibinfo {author} {\bibfnamefont {S.}~\bibnamefont {Sugiyama}}, \bibinfo
  {author} {\bibfnamefont {T.}~\bibnamefont {Mikado}}, \bibinfo {author}
  {\bibfnamefont {M.}~\bibnamefont {Chiwaki}}, \bibinfo {author} {\bibfnamefont
  {K.}~\bibnamefont {Yamada}}, \bibinfo {author} {\bibfnamefont
  {R.}~\bibnamefont {Suzuki}}, \bibinfo {author} {\bibfnamefont
  {N.}~\bibnamefont {Sei}}, \bibinfo {author} {\bibfnamefont {T.}~\bibnamefont
  {Ohdaira}},\ and\ \bibinfo {author} {\bibfnamefont {T.}~\bibnamefont
  {Yamazaki}},\ }\bibfield  {title} {\bibinfo {title} {Polarized gamma-rays
  with laser-compton backscattering},\ }\href
  {https://doi.org/https://doi.org/10.1016/0168-9002(95)01216-8} {\bibfield
  {journal} {\bibinfo  {journal} {Nuclear Instruments and Methods in Physics
  Research Section A: Accelerators, Spectrometers, Detectors and Associated
  Equipment}\ }\textbf {\bibinfo {volume} {375}},\ \bibinfo {pages} {602}
  (\bibinfo {year} {1996})},\ \bibinfo {note} {proceedings of the 17th
  International Free Electron Laser Conference}\BibitemShut {NoStop}%
\bibitem [{\citenamefont {Omori}\ \emph {et~al.}(2006)\citenamefont {Omori},
  \citenamefont {Fukuda}, \citenamefont {Hirose}, \citenamefont {Kurihara},
  \citenamefont {Kuroda}, \citenamefont {Nomura}, \citenamefont {Ohashi},
  \citenamefont {Okugi}, \citenamefont {Sakaue}, \citenamefont {Saito},
  \citenamefont {Urakawa}, \citenamefont {Washio},\ and\ \citenamefont
  {Yamazaki}}]{Omori_2006}%
  \BibitemOpen
  \bibfield  {author} {\bibinfo {author} {\bibfnamefont {T.}~\bibnamefont
  {Omori}}, \bibinfo {author} {\bibfnamefont {M.}~\bibnamefont {Fukuda}},
  \bibinfo {author} {\bibfnamefont {T.}~\bibnamefont {Hirose}}, \bibinfo
  {author} {\bibfnamefont {Y.}~\bibnamefont {Kurihara}}, \bibinfo {author}
  {\bibfnamefont {R.}~\bibnamefont {Kuroda}}, \bibinfo {author} {\bibfnamefont
  {M.}~\bibnamefont {Nomura}}, \bibinfo {author} {\bibfnamefont
  {A.}~\bibnamefont {Ohashi}}, \bibinfo {author} {\bibfnamefont
  {T.}~\bibnamefont {Okugi}}, \bibinfo {author} {\bibfnamefont
  {K.}~\bibnamefont {Sakaue}}, \bibinfo {author} {\bibfnamefont
  {T.}~\bibnamefont {Saito}}, \bibinfo {author} {\bibfnamefont
  {J.}~\bibnamefont {Urakawa}}, \bibinfo {author} {\bibfnamefont
  {M.}~\bibnamefont {Washio}},\ and\ \bibinfo {author} {\bibfnamefont
  {I.}~\bibnamefont {Yamazaki}},\ }\bibfield  {title} {\bibinfo {title}
  {Efficient propagation of polarization from laser photons to positrons
  through compton scattering and electron-positron pair creation},\ }\href
  {https://doi.org/10.1103/PhysRevLett.96.114801} {\bibfield  {journal}
  {\bibinfo  {journal} {Phys. Rev. Lett.}\ }\textbf {\bibinfo {volume} {96}},\
  \bibinfo {pages} {114801} (\bibinfo {year} {2006})}\BibitemShut {NoStop}%
\bibitem [{\citenamefont {Seipt}\ \emph {et~al.}(2018)\citenamefont {Seipt},
  \citenamefont {Del~Sorbo}, \citenamefont {Ridgers},\ and\ \citenamefont
  {Thomas}}]{Seipt2018}%
  \BibitemOpen
  \bibfield  {author} {\bibinfo {author} {\bibfnamefont {D.}~\bibnamefont
  {Seipt}}, \bibinfo {author} {\bibfnamefont {D.}~\bibnamefont {Del~Sorbo}},
  \bibinfo {author} {\bibfnamefont {C.~P.}\ \bibnamefont {Ridgers}},\ and\
  \bibinfo {author} {\bibfnamefont {A.~G.~R.}\ \bibnamefont {Thomas}},\
  }\bibfield  {title} {\bibinfo {title} {Theory of radiative electron
  polarization in strong laser fields},\ }\href
  {https://doi.org/10.1103/PhysRevA.98.023417} {\bibfield  {journal} {\bibinfo
  {journal} {Phys. Rev. A}\ }\textbf {\bibinfo {volume} {98}},\ \bibinfo
  {pages} {023417} (\bibinfo {year} {2018})}\BibitemShut {NoStop}%
\bibitem [{\citenamefont {Kotkin}\ \emph {et~al.}(2003)\citenamefont {Kotkin},
  \citenamefont {Serbo},\ and\ \citenamefont {Telnov}}]{Kotkin2003}%
  \BibitemOpen
  \bibfield  {author} {\bibinfo {author} {\bibfnamefont {G.~L.}\ \bibnamefont
  {Kotkin}}, \bibinfo {author} {\bibfnamefont {V.~G.}\ \bibnamefont {Serbo}},\
  and\ \bibinfo {author} {\bibfnamefont {V.~I.}\ \bibnamefont {Telnov}},\
  }\bibfield  {title} {\bibinfo {title} {Electron (positron) beam polarization
  by compton scattering on circularly polarized laser photons},\ }\href
  {https://doi.org/10.1103/PhysRevSTAB.6.011001} {\bibfield  {journal}
  {\bibinfo  {journal} {Phys. Rev. ST Accel. Beams}\ }\textbf {\bibinfo
  {volume} {6}},\ \bibinfo {pages} {011001} (\bibinfo {year}
  {2003})}\BibitemShut {NoStop}%
\bibitem [{\citenamefont {Karlovets}(2011)}]{Karlovets2011}%
  \BibitemOpen
  \bibfield  {author} {\bibinfo {author} {\bibfnamefont {D.~V.}\ \bibnamefont
  {Karlovets}},\ }\bibfield  {title} {\bibinfo {title} {Radiative polarization
  of electrons in a strong laser wave},\ }\href
  {https://doi.org/10.1103/PhysRevA.84.062116} {\bibfield  {journal} {\bibinfo
  {journal} {Phys. Rev. A}\ }\textbf {\bibinfo {volume} {84}},\ \bibinfo
  {pages} {062116} (\bibinfo {year} {2011})}\BibitemShut {NoStop}%
\bibitem [{\citenamefont {Di~Piazza}\ \emph {et~al.}(2020)\citenamefont
  {Di~Piazza}, \citenamefont {Wistisen}, \citenamefont {Tamburini},\ and\
  \citenamefont {Uggerh{\o}j}}]{di2020testing}%
  \BibitemOpen
  \bibfield  {author} {\bibinfo {author} {\bibfnamefont {A.}~\bibnamefont
  {Di~Piazza}}, \bibinfo {author} {\bibfnamefont {T.}~\bibnamefont {Wistisen}},
  \bibinfo {author} {\bibfnamefont {M.}~\bibnamefont {Tamburini}},\ and\
  \bibinfo {author} {\bibfnamefont {U.}~\bibnamefont {Uggerh{\o}j}},\
  }\bibfield  {title} {\bibinfo {title} {Testing strong field qed close to the
  fully nonperturbative regime using aligned crystals},\ }\href@noop {}
  {\bibfield  {journal} {\bibinfo  {journal} {Physical review letters}\
  }\textbf {\bibinfo {volume} {124}},\ \bibinfo {pages} {044801} (\bibinfo
  {year} {2020})}\BibitemShut {NoStop}%
\bibitem [{\citenamefont {Blackburn}\ \emph {et~al.}(2019)\citenamefont
  {Blackburn}, \citenamefont {Ilderton}, \citenamefont {Marklund},\ and\
  \citenamefont {Ridgers}}]{blackburn2019reaching}%
  \BibitemOpen
  \bibfield  {author} {\bibinfo {author} {\bibfnamefont {T.}~\bibnamefont
  {Blackburn}}, \bibinfo {author} {\bibfnamefont {A.}~\bibnamefont {Ilderton}},
  \bibinfo {author} {\bibfnamefont {M.}~\bibnamefont {Marklund}},\ and\
  \bibinfo {author} {\bibfnamefont {C.}~\bibnamefont {Ridgers}},\ }\bibfield
  {title} {\bibinfo {title} {Reaching supercritical field strengths with
  intense lasers},\ }\href@noop {} {\bibfield  {journal} {\bibinfo  {journal}
  {New Journal of Physics}\ }\textbf {\bibinfo {volume} {21}},\ \bibinfo
  {pages} {053040} (\bibinfo {year} {2019})}\BibitemShut {NoStop}%
\bibitem [{\citenamefont {Baumann}\ \emph {et~al.}(2019)\citenamefont
  {Baumann}, \citenamefont {Nerush}, \citenamefont {Pukhov},\ and\
  \citenamefont {Kostyukov}}]{baumann2019probing}%
  \BibitemOpen
  \bibfield  {author} {\bibinfo {author} {\bibfnamefont {C.}~\bibnamefont
  {Baumann}}, \bibinfo {author} {\bibfnamefont {E.}~\bibnamefont {Nerush}},
  \bibinfo {author} {\bibfnamefont {A.}~\bibnamefont {Pukhov}},\ and\ \bibinfo
  {author} {\bibfnamefont {I.~Y.}\ \bibnamefont {Kostyukov}},\ }\bibfield
  {title} {\bibinfo {title} {Probing non-perturbative qed with electron-laser
  collisions},\ }\href@noop {} {\bibfield  {journal} {\bibinfo  {journal}
  {Scientific reports}\ }\textbf {\bibinfo {volume} {9}},\ \bibinfo {pages} {1}
  (\bibinfo {year} {2019})}\BibitemShut {NoStop}%
\bibitem [{\citenamefont {Fedotov}(2017)}]{fedotov2017conjecture}%
  \BibitemOpen
  \bibfield  {author} {\bibinfo {author} {\bibfnamefont {A.}~\bibnamefont
  {Fedotov}},\ }\bibfield  {title} {\bibinfo {title} {Conjecture of
  perturbative qed breakdown at $\alpha$$\chi$2/3$\gtrsim$1},\ }in\ \href@noop
  {} {\emph {\bibinfo {booktitle} {Journal of Physics: Conference Series}}},\
  Vol.\ \bibinfo {volume} {826}\ (\bibinfo {organization} {IOP Publishing},\
  \bibinfo {year} {2017})\ p.\ \bibinfo {pages} {012027}\BibitemShut {NoStop}%
\bibitem [{\citenamefont {Yakimenko}\ \emph {et~al.}(2019)\citenamefont
  {Yakimenko}, \citenamefont {Meuren}, \citenamefont {Del~Gaudio},
  \citenamefont {Baumann}, \citenamefont {Fedotov}, \citenamefont {Fiuza},
  \citenamefont {Grismayer}, \citenamefont {Hogan}, \citenamefont {Pukhov},
  \citenamefont {Silva} \emph {et~al.}}]{yakimenko2019prospect}%
  \BibitemOpen
  \bibfield  {author} {\bibinfo {author} {\bibfnamefont {V.}~\bibnamefont
  {Yakimenko}}, \bibinfo {author} {\bibfnamefont {S.}~\bibnamefont {Meuren}},
  \bibinfo {author} {\bibfnamefont {F.}~\bibnamefont {Del~Gaudio}}, \bibinfo
  {author} {\bibfnamefont {C.}~\bibnamefont {Baumann}}, \bibinfo {author}
  {\bibfnamefont {A.}~\bibnamefont {Fedotov}}, \bibinfo {author} {\bibfnamefont
  {F.}~\bibnamefont {Fiuza}}, \bibinfo {author} {\bibfnamefont
  {T.}~\bibnamefont {Grismayer}}, \bibinfo {author} {\bibfnamefont
  {M.}~\bibnamefont {Hogan}}, \bibinfo {author} {\bibfnamefont
  {A.}~\bibnamefont {Pukhov}}, \bibinfo {author} {\bibfnamefont
  {L.}~\bibnamefont {Silva}}, \emph {et~al.},\ }\bibfield  {title} {\bibinfo
  {title} {Prospect of studying nonperturbative qed with beam-beam
  collisions},\ }\href@noop {} {\bibfield  {journal} {\bibinfo  {journal}
  {Physical review letters}\ }\textbf {\bibinfo {volume} {122}},\ \bibinfo
  {pages} {190404} (\bibinfo {year} {2019})}\BibitemShut {NoStop}%
\bibitem [{\citenamefont {Mironov}\ \emph {et~al.}(2020)\citenamefont
  {Mironov}, \citenamefont {Meuren},\ and\ \citenamefont
  {Fedotov}}]{Mironov2020}%
  \BibitemOpen
  \bibfield  {author} {\bibinfo {author} {\bibfnamefont {A.~A.}\ \bibnamefont
  {Mironov}}, \bibinfo {author} {\bibfnamefont {S.}~\bibnamefont {Meuren}},\
  and\ \bibinfo {author} {\bibfnamefont {A.~M.}\ \bibnamefont {Fedotov}},\
  }\bibfield  {title} {\bibinfo {title} {Resummation of qed radiative
  corrections in a strong constant crossed field},\ }\href
  {https://doi.org/10.1103/PhysRevD.102.053005} {\bibfield  {journal} {\bibinfo
   {journal} {Phys. Rev. D}\ }\textbf {\bibinfo {volume} {102}},\ \bibinfo
  {pages} {053005} (\bibinfo {year} {2020})}\BibitemShut {NoStop}%
\bibitem [{\citenamefont {Ritus}(1970)}]{ritus1970radiative}%
  \BibitemOpen
  \bibfield  {author} {\bibinfo {author} {\bibfnamefont {V.}~\bibnamefont
  {Ritus}},\ }\bibfield  {title} {\bibinfo {title} {Radiative effects and their
  enhancement in an intense electromagnetic field},\ }\href@noop {} {\bibfield
  {journal} {\bibinfo  {journal} {Sov. Phys. JETP}\ }\textbf {\bibinfo {volume}
  {30}},\ \bibinfo {pages} {052805} (\bibinfo {year} {1970})}\BibitemShut
  {NoStop}%
\bibitem [{\citenamefont {Hanneke}\ \emph {et~al.}(2008)\citenamefont
  {Hanneke}, \citenamefont {Fogwell},\ and\ \citenamefont
  {Gabrielse}}]{Hanneke2008}%
  \BibitemOpen
  \bibfield  {author} {\bibinfo {author} {\bibfnamefont {D.}~\bibnamefont
  {Hanneke}}, \bibinfo {author} {\bibfnamefont {S.}~\bibnamefont {Fogwell}},\
  and\ \bibinfo {author} {\bibfnamefont {G.}~\bibnamefont {Gabrielse}},\
  }\bibfield  {title} {\bibinfo {title} {New measurement of the electron
  magnetic moment and the fine structure constant},\ }\href
  {https://doi.org/10.1103/PhysRevLett.100.120801} {\bibfield  {journal}
  {\bibinfo  {journal} {Phys. Rev. Lett.}\ }\textbf {\bibinfo {volume} {100}},\
  \bibinfo {pages} {120801} (\bibinfo {year} {2008})}\BibitemShut {NoStop}%
\bibitem [{\citenamefont {Sturm}\ \emph {et~al.}(2011)\citenamefont {Sturm},
  \citenamefont {Wagner}, \citenamefont {Schabinger}, \citenamefont {Zatorski},
  \citenamefont {Harman}, \citenamefont {Quint}, \citenamefont {Werth},
  \citenamefont {Keitel},\ and\ \citenamefont {Blaum}}]{Sturm2011}%
  \BibitemOpen
  \bibfield  {author} {\bibinfo {author} {\bibfnamefont {S.}~\bibnamefont
  {Sturm}}, \bibinfo {author} {\bibfnamefont {A.}~\bibnamefont {Wagner}},
  \bibinfo {author} {\bibfnamefont {B.}~\bibnamefont {Schabinger}}, \bibinfo
  {author} {\bibfnamefont {J.}~\bibnamefont {Zatorski}}, \bibinfo {author}
  {\bibfnamefont {Z.}~\bibnamefont {Harman}}, \bibinfo {author} {\bibfnamefont
  {W.}~\bibnamefont {Quint}}, \bibinfo {author} {\bibfnamefont
  {G.}~\bibnamefont {Werth}}, \bibinfo {author} {\bibfnamefont {C.~H.}\
  \bibnamefont {Keitel}},\ and\ \bibinfo {author} {\bibfnamefont
  {K.}~\bibnamefont {Blaum}},\ }\bibfield  {title} {\bibinfo {title} {$g$
  factor of hydrogenlike $^{28}\mathrm{Si}^{13+}$},\ }\href
  {https://doi.org/10.1103/PhysRevLett.107.023002} {\bibfield  {journal}
  {\bibinfo  {journal} {Phys. Rev. Lett.}\ }\textbf {\bibinfo {volume} {107}},\
  \bibinfo {pages} {023002} (\bibinfo {year} {2011})}\BibitemShut {NoStop}%
\bibitem [{\citenamefont {Abi}\ \emph {et~al.}(2021)\citenamefont {Abi},
  \citenamefont {Albahri}, \citenamefont {Al-Kilani},\ and\ \citenamefont
  {\emph{et al.}}}]{Abi2021}%
  \BibitemOpen
  \bibfield  {author} {\bibinfo {author} {\bibfnamefont {B.}~\bibnamefont
  {Abi}}, \bibinfo {author} {\bibfnamefont {T.}~\bibnamefont {Albahri}},
  \bibinfo {author} {\bibfnamefont {S.}~\bibnamefont {Al-Kilani}},\ and\
  \bibinfo {author} {\bibnamefont {\emph{et al.}}} (\bibinfo {collaboration}
  {Muon $g\ensuremath{-}2$ Collaboration}),\ }\bibfield  {title} {\bibinfo
  {title} {Measurement of the positive muon anomalous magnetic moment to 0.46
  ppm},\ }\href {https://doi.org/10.1103/PhysRevLett.126.141801} {\bibfield
  {journal} {\bibinfo  {journal} {Phys. Rev. Lett.}\ }\textbf {\bibinfo
  {volume} {126}},\ \bibinfo {pages} {141801} (\bibinfo {year}
  {2021})}\BibitemShut {NoStop}%
\bibitem [{\citenamefont {Ridgers}\ \emph {et~al.}(2014)\citenamefont
  {Ridgers}, \citenamefont {Kirk}, \citenamefont {Duclous}, \citenamefont
  {Blackburn}, \citenamefont {Brady}, \citenamefont {Bennett}, \citenamefont
  {Arber},\ and\ \citenamefont {Bell}}]{Ridgers2014}%
  \BibitemOpen
  \bibfield  {author} {\bibinfo {author} {\bibfnamefont {C.}~\bibnamefont
  {Ridgers}}, \bibinfo {author} {\bibfnamefont {J.}~\bibnamefont {Kirk}},
  \bibinfo {author} {\bibfnamefont {R.}~\bibnamefont {Duclous}}, \bibinfo
  {author} {\bibfnamefont {T.}~\bibnamefont {Blackburn}}, \bibinfo {author}
  {\bibfnamefont {C.}~\bibnamefont {Brady}}, \bibinfo {author} {\bibfnamefont
  {K.}~\bibnamefont {Bennett}}, \bibinfo {author} {\bibfnamefont
  {T.}~\bibnamefont {Arber}},\ and\ \bibinfo {author} {\bibfnamefont
  {A.}~\bibnamefont {Bell}},\ }\bibfield  {title} {\bibinfo {title} {Modelling
  gamma-ray photon emission and pair production in high-intensity
  laser–matter interactions},\ }\href
  {https://doi.org/https://doi.org/10.1016/j.jcp.2013.12.007} {\bibfield
  {journal} {\bibinfo  {journal} {Journal of Computational Physics}\ }\textbf
  {\bibinfo {volume} {260}},\ \bibinfo {pages} {273 } (\bibinfo {year}
  {2014})}\BibitemShut {NoStop}%
\bibitem [{\citenamefont {Elkina}\ \emph {et~al.}(2011)\citenamefont {Elkina},
  \citenamefont {Fedotov}, \citenamefont {Kostyukov}, \citenamefont {Legkov},
  \citenamefont {Narozhny}, \citenamefont {Nerush},\ and\ \citenamefont
  {Ruhl}}]{Elkina2011}%
  \BibitemOpen
  \bibfield  {author} {\bibinfo {author} {\bibfnamefont {N.~V.}\ \bibnamefont
  {Elkina}}, \bibinfo {author} {\bibfnamefont {A.~M.}\ \bibnamefont {Fedotov}},
  \bibinfo {author} {\bibfnamefont {I.~Y.}\ \bibnamefont {Kostyukov}}, \bibinfo
  {author} {\bibfnamefont {M.~V.}\ \bibnamefont {Legkov}}, \bibinfo {author}
  {\bibfnamefont {N.~B.}\ \bibnamefont {Narozhny}}, \bibinfo {author}
  {\bibfnamefont {E.~N.}\ \bibnamefont {Nerush}},\ and\ \bibinfo {author}
  {\bibfnamefont {H.}~\bibnamefont {Ruhl}},\ }\bibfield  {title} {\bibinfo
  {title} {Qed cascades induced by circularly polarized laser fields},\ }\href
  {https://doi.org/10.1103/PhysRevSTAB.14.054401} {\bibfield  {journal}
  {\bibinfo  {journal} {Phys. Rev. ST Accel. Beams}\ }\textbf {\bibinfo
  {volume} {14}},\ \bibinfo {pages} {054401} (\bibinfo {year}
  {2011})}\BibitemShut {NoStop}%
\bibitem [{\citenamefont {Gonoskov}\ \emph {et~al.}(2015)\citenamefont
  {Gonoskov}, \citenamefont {Bastrakov}, \citenamefont {Efimenko},
  \citenamefont {Ilderton}, \citenamefont {Marklund}, \citenamefont {Meyerov},
  \citenamefont {Muraviev}, \citenamefont {Sergeev}, \citenamefont {Surmin},\
  and\ \citenamefont {Wallin}}]{Gonoskov2015}%
  \BibitemOpen
  \bibfield  {author} {\bibinfo {author} {\bibfnamefont {A.}~\bibnamefont
  {Gonoskov}}, \bibinfo {author} {\bibfnamefont {S.}~\bibnamefont {Bastrakov}},
  \bibinfo {author} {\bibfnamefont {E.}~\bibnamefont {Efimenko}}, \bibinfo
  {author} {\bibfnamefont {A.}~\bibnamefont {Ilderton}}, \bibinfo {author}
  {\bibfnamefont {M.}~\bibnamefont {Marklund}}, \bibinfo {author}
  {\bibfnamefont {I.}~\bibnamefont {Meyerov}}, \bibinfo {author} {\bibfnamefont
  {A.}~\bibnamefont {Muraviev}}, \bibinfo {author} {\bibfnamefont
  {A.}~\bibnamefont {Sergeev}}, \bibinfo {author} {\bibfnamefont
  {I.}~\bibnamefont {Surmin}},\ and\ \bibinfo {author} {\bibfnamefont
  {E.}~\bibnamefont {Wallin}},\ }\bibfield  {title} {\bibinfo {title} {Extended
  particle-in-cell schemes for physics in ultrastrong laser fields: Review and
  developments},\ }\href {https://doi.org/10.1103/PhysRevE.92.023305}
  {\bibfield  {journal} {\bibinfo  {journal} {Phys. Rev. E}\ }\textbf {\bibinfo
  {volume} {92}},\ \bibinfo {pages} {023305} (\bibinfo {year}
  {2015})}\BibitemShut {NoStop}%
\bibitem [{\citenamefont {of~CAIN Version~2.42}()}]{CAIN}%
  \BibitemOpen
  \bibfield  {author} {\bibinfo {author} {\bibfnamefont {U.~M.}\ \bibnamefont
  {of~CAIN Version~2.42}},\ }\href@noop {} {\ }\bibinfo {note} {\url{
  http://lcdev.kek.jp/~yokoya/CAIN/}}\BibitemShut {NoStop}%
\bibitem [{\citenamefont {Ritus}(1985)}]{Ritus1985}%
  \BibitemOpen
  \bibfield  {author} {\bibinfo {author} {\bibfnamefont {V.~I.}\ \bibnamefont
  {Ritus}},\ }\href@noop {} {\bibfield  {journal} {\bibinfo  {journal} {J. Sov.
  Laser Res.}\ }\textbf {\bibinfo {volume} {6}},\ \bibinfo {pages} {497}
  (\bibinfo {year} {1985})}\BibitemShut {NoStop}%
\bibitem [{\citenamefont {Di~Piazza}\ \emph {et~al.}(2018)\citenamefont
  {Di~Piazza}, \citenamefont {Tamburini}, \citenamefont {Meuren},\ and\
  \citenamefont {Keitel}}]{Piazza2018}%
  \BibitemOpen
  \bibfield  {author} {\bibinfo {author} {\bibfnamefont {A.}~\bibnamefont
  {Di~Piazza}}, \bibinfo {author} {\bibfnamefont {M.}~\bibnamefont
  {Tamburini}}, \bibinfo {author} {\bibfnamefont {S.}~\bibnamefont {Meuren}},\
  and\ \bibinfo {author} {\bibfnamefont {C.~H.}\ \bibnamefont {Keitel}},\
  }\bibfield  {title} {\bibinfo {title} {Implementing nonlinear compton
  scattering beyond the local-constant-field approximation},\ }\href
  {https://doi.org/10.1103/PhysRevA.98.012134} {\bibfield  {journal} {\bibinfo
  {journal} {Phys. Rev. A}\ }\textbf {\bibinfo {volume} {98}},\ \bibinfo
  {pages} {012134} (\bibinfo {year} {2018})}\BibitemShut {NoStop}%
\bibitem [{\citenamefont {Di~Piazza}\ \emph {et~al.}(2019)\citenamefont
  {Di~Piazza}, \citenamefont {Tamburini}, \citenamefont {Meuren},\ and\
  \citenamefont {Keitel}}]{Piazza2019}%
  \BibitemOpen
  \bibfield  {author} {\bibinfo {author} {\bibfnamefont {A.}~\bibnamefont
  {Di~Piazza}}, \bibinfo {author} {\bibfnamefont {M.}~\bibnamefont
  {Tamburini}}, \bibinfo {author} {\bibfnamefont {S.}~\bibnamefont {Meuren}},\
  and\ \bibinfo {author} {\bibfnamefont {C.~H.}\ \bibnamefont {Keitel}},\
  }\bibfield  {title} {\bibinfo {title} {Improved local-constant-field
  approximation for strong-field qed codes},\ }\href
  {https://doi.org/10.1103/PhysRevA.99.022125} {\bibfield  {journal} {\bibinfo
  {journal} {Phys. Rev. A}\ }\textbf {\bibinfo {volume} {99}},\ \bibinfo
  {pages} {022125} (\bibinfo {year} {2019})}\BibitemShut {NoStop}%
\bibitem [{\citenamefont {Ilderton}\ \emph {et~al.}(2019)\citenamefont
  {Ilderton}, \citenamefont {King},\ and\ \citenamefont
  {Seipt}}]{Ilderton2019}%
  \BibitemOpen
  \bibfield  {author} {\bibinfo {author} {\bibfnamefont {A.}~\bibnamefont
  {Ilderton}}, \bibinfo {author} {\bibfnamefont {B.}~\bibnamefont {King}},\
  and\ \bibinfo {author} {\bibfnamefont {D.}~\bibnamefont {Seipt}},\ }\bibfield
   {title} {\bibinfo {title} {Extended locally constant field approximation for
  nonlinear compton scattering},\ }\href
  {https://doi.org/10.1103/PhysRevA.99.042121} {\bibfield  {journal} {\bibinfo
  {journal} {Phys. Rev. A}\ }\textbf {\bibinfo {volume} {99}},\ \bibinfo
  {pages} {042121} (\bibinfo {year} {2019})}\BibitemShut {NoStop}%
\bibitem [{\citenamefont {Lv}\ \emph {et~al.}(2021)\citenamefont {Lv},
  \citenamefont {Raicher}, \citenamefont {Keitel},\ and\ \citenamefont
  {Hatsagortsyan}}]{lv2021anomalous}%
  \BibitemOpen
  \bibfield  {author} {\bibinfo {author} {\bibfnamefont {Q.}~\bibnamefont
  {Lv}}, \bibinfo {author} {\bibfnamefont {E.}~\bibnamefont {Raicher}},
  \bibinfo {author} {\bibfnamefont {C.}~\bibnamefont {Keitel}},\ and\ \bibinfo
  {author} {\bibfnamefont {K.}~\bibnamefont {Hatsagortsyan}},\ }\bibfield
  {title} {\bibinfo {title} {Anomalous violation of the local constant field
  approximation in colliding laser beams},\ }\href@noop {} {\bibfield
  {journal} {\bibinfo  {journal} {Physical Review Research}\ }\textbf {\bibinfo
  {volume} {3}},\ \bibinfo {pages} {013214} (\bibinfo {year}
  {2021})}\BibitemShut {NoStop}%
\bibitem [{SM()}]{SM}%
  \BibitemOpen
  \href@noop {} {}\bibinfo {note} {See Supplemental Materials for details on
  the applied theoretical model, simulated results for other laser and electron
  parameters, one-loop effects, and impact of laser polarization.}\BibitemShut
  {Stop}%
\bibitem [{\citenamefont {Tang}\ \emph {et~al.}(2021)\citenamefont {Tang},
  \citenamefont {Gong}, \citenamefont {Yu}, \citenamefont {Shou},\ and\
  \citenamefont {Yan}}]{Tang2021}%
  \BibitemOpen
  \bibfield  {author} {\bibinfo {author} {\bibfnamefont {Y.}~\bibnamefont
  {Tang}}, \bibinfo {author} {\bibfnamefont {Z.}~\bibnamefont {Gong}}, \bibinfo
  {author} {\bibfnamefont {J.}~\bibnamefont {Yu}}, \bibinfo {author}
  {\bibfnamefont {Y.}~\bibnamefont {Shou}},\ and\ \bibinfo {author}
  {\bibfnamefont {X.}~\bibnamefont {Yan}},\ }\bibfield  {title} {\bibinfo
  {title} {Radiative polarization dynamics of relativistic electrons in an
  intense electromagnetic field},\ }\href
  {https://doi.org/10.1103/PhysRevA.103.042807} {\bibfield  {journal} {\bibinfo
   {journal} {Phys. Rev. A}\ }\textbf {\bibinfo {volume} {103}},\ \bibinfo
  {pages} {042807} (\bibinfo {year} {2021})}\BibitemShut {NoStop}%
\bibitem [{\citenamefont {Kotkin}\ \emph {et~al.}(1998)\citenamefont {Kotkin},
  \citenamefont {Perlt},\ and\ \citenamefont {Serbo}}]{kotkin1998polarization}%
  \BibitemOpen
  \bibfield  {author} {\bibinfo {author} {\bibfnamefont {G.}~\bibnamefont
  {Kotkin}}, \bibinfo {author} {\bibfnamefont {H.}~\bibnamefont {Perlt}},\ and\
  \bibinfo {author} {\bibfnamefont {V.}~\bibnamefont {Serbo}},\ }\bibfield
  {title} {\bibinfo {title} {Polarization of high-energy electrons traversing a
  laser beam},\ }\href@noop {} {\bibfield  {journal} {\bibinfo  {journal}
  {Nuclear Instruments and Methods in Physics Research Section A: Accelerators,
  Spectrometers, Detectors and Associated Equipment}\ }\textbf {\bibinfo
  {volume} {404}},\ \bibinfo {pages} {430} (\bibinfo {year}
  {1998})}\BibitemShut {NoStop}%
\bibitem [{\citenamefont {Bargmann}\ \emph {et~al.}(1959)\citenamefont
  {Bargmann}, \citenamefont {Michel},\ and\ \citenamefont
  {Telegdi}}]{Bargmann1959}%
  \BibitemOpen
  \bibfield  {author} {\bibinfo {author} {\bibfnamefont {V.}~\bibnamefont
  {Bargmann}}, \bibinfo {author} {\bibfnamefont {L.}~\bibnamefont {Michel}},\
  and\ \bibinfo {author} {\bibfnamefont {V.~L.}\ \bibnamefont {Telegdi}},\
  }\bibfield  {title} {\bibinfo {title} {Precession of the polarization of
  particles moving in a homogeneous electromagnetic field},\ }\href
  {https://doi.org/10.1103/PhysRevLett.2.435} {\bibfield  {journal} {\bibinfo
  {journal} {Phys. Rev. Lett.}\ }\textbf {\bibinfo {volume} {2}},\ \bibinfo
  {pages} {435} (\bibinfo {year} {1959})}\BibitemShut {NoStop}%
\bibitem [{\citenamefont {Wen}\ \emph {et~al.}(2016)\citenamefont {Wen},
  \citenamefont {Bauke},\ and\ \citenamefont {Keitel}}]{Wen_2016}%
  \BibitemOpen
  \bibfield  {author} {\bibinfo {author} {\bibfnamefont {M.}~\bibnamefont
  {Wen}}, \bibinfo {author} {\bibfnamefont {H.}~\bibnamefont {Bauke}},\ and\
  \bibinfo {author} {\bibfnamefont {C.~H.}\ \bibnamefont {Keitel}},\ }\bibfield
   {title} {\bibinfo {title} {Identifying the stern-gerlach force of classical
  electron dynamics},\ }\href@noop {} {\bibfield  {journal} {\bibinfo
  {journal} {Scientific Reports}\ }\textbf {\bibinfo {volume} {6}},\ \bibinfo
  {pages} {31624} (\bibinfo {year} {2016})}\BibitemShut {NoStop}%
\bibitem [{\citenamefont {Formanek}\ \emph {et~al.}(2021)\citenamefont
  {Formanek}, \citenamefont {Steinmetz},\ and\ \citenamefont
  {Rafelski}}]{Formanek_2021}%
  \BibitemOpen
  \bibfield  {author} {\bibinfo {author} {\bibfnamefont {M.}~\bibnamefont
  {Formanek}}, \bibinfo {author} {\bibfnamefont {A.}~\bibnamefont
  {Steinmetz}},\ and\ \bibinfo {author} {\bibfnamefont {J.}~\bibnamefont
  {Rafelski}},\ }\bibfield  {title} {\bibinfo {title} {Motion of classical
  charged particles with magnetic moment in external plane-wave electromagnetic
  fields},\ }\href {https://doi.org/10.1103/PhysRevA.103.052218} {\bibfield
  {journal} {\bibinfo  {journal} {Phys. Rev. A}\ }\textbf {\bibinfo {volume}
  {103}},\ \bibinfo {pages} {052218} (\bibinfo {year} {2021})}\BibitemShut
  {NoStop}%
\bibitem [{SM2()}]{SM2}%
  \BibitemOpen
  \href@noop {} {}\bibinfo {note} {See Sec. VI of Supplemental Materials for
  the justification of the simplifications.}\BibitemShut {Stop}%
\bibitem [{\citenamefont {Jackson}(1999{\natexlab{b}})}]{jackson1999classical}%
  \BibitemOpen
  \bibfield  {author} {\bibinfo {author} {\bibfnamefont {J.~D.}\ \bibnamefont
  {Jackson}},\ }\href@noop {} {\emph {\bibinfo {title} {Classical
  electrodynamics}}}\ (\bibinfo  {publisher} {American Association of Physics
  Teachers},\ \bibinfo {year} {1999})\BibitemShut {NoStop}%
\bibitem [{\citenamefont {Rich}\ and\ \citenamefont
  {Wesley}(1972)}]{rich1972current}%
  \BibitemOpen
  \bibfield  {author} {\bibinfo {author} {\bibfnamefont {A.}~\bibnamefont
  {Rich}}\ and\ \bibinfo {author} {\bibfnamefont {J.~C.}\ \bibnamefont
  {Wesley}},\ }\bibfield  {title} {\bibinfo {title} {The current status of the
  lepton g factors},\ }\href@noop {} {\bibfield  {journal} {\bibinfo  {journal}
  {Reviews of Modern Physics}\ }\textbf {\bibinfo {volume} {44}},\ \bibinfo
  {pages} {250} (\bibinfo {year} {1972})}\BibitemShut {NoStop}%
\end{thebibliography}%

\end{document}